\documentclass[12pt,preprint]{aastex}
\slugcomment{Submitted to {\it The Astrophysical Journal}}
\shortauthors{Kirkpatrick}
\shorttitle{WISE Y Dwarfs}

\usepackage{color}

\begin{document}

\title{Further Defining Spectral Type ``Y'' and Exploring the Low-mass End of the Field Brown Dwarf Mass Function}

\author{J.\ Davy Kirkpatrick\altaffilmark{a},
Christopher R.\ Gelino\altaffilmark{a},
Michael C.\ Cushing\altaffilmark{b},
Gregory N.\ Mace\altaffilmark{c}
Roger L.\ Griffith\altaffilmark{a},
Michael F.\ Skrutskie\altaffilmark{d},
Kenneth A.\ Marsh\altaffilmark{a},
Edward L.\ Wright\altaffilmark{c},
Peter R.\ Eisenhardt\altaffilmark{e},
Ian S.\ McLean\altaffilmark{c},
Amanda K.\ Mainzer\altaffilmark{e},
Adam J.\ Burgasser\altaffilmark{f},
C.\ G.\ Tinney\altaffilmark{g},
Stephen Parker\altaffilmark{g},
Graeme Salter\altaffilmark{g}
}

\altaffiltext{a}{Infrared Processing and Analysis Center, MS 100-22, California 
    Institute of Technology, Pasadena, CA 91125; davy@ipac.caltech.edu}
\altaffiltext{b}{Department of Physics and Astronomy, MS 111, University of Toledo, 2801 W. Bancroft St., Toledo, OH 43606-3328}
\altaffiltext{c}{Department of Physics and Astronomy, UCLA, Los Angeles, CA 90095-1547}
\altaffiltext{d}{Department of Astronomy, University of Virginia, Charlottesville, VA, 22904}
\altaffiltext{e}{NASA Jet Propulsion Laboratory, 4800 Oak Grove Drive, Pasadena, CA 91109}
\altaffiltext{f}{Department of Physics, University of California, San Diego, CA 92093}
\altaffiltext{g}{Department of Astrophysics, School of Physics, University of New South Wales, NSW 2052, Australia}

\begin{abstract}

We present the discovery of another seven Y dwarfs from the Wide-field Infrared Survey Explorer (WISE).
Using these objects, as well as the first six WISE Y dwarf discoveries from Cushing et al., we further
explore the transition between spectral types T and Y. We find that the T/Y boundary roughly coincides
with the spot where the $J-H$ colors of brown dwarfs, as predicted by models, turn
back to the red. Moreover, we use preliminary trigonometric parallax measurements 
to show that the T/Y boundary may also correspond to the point at which the
absolute $H$ (1.6 $\mu$m) and W2 (4.6 $\mu$m) magnitudes plummet. We use these discoveries and their preliminary
distances to place them in the larger context of the Solar Neighborhood. We present a table that
updates the entire stellar and substellar constinuency within 8 parsecs of the Sun, and we show that
the current census has hydrogen-burning stars outnumbering brown dwarfs by roughly a factor of six. 
This factor will decrease with time as more brown dwarfs are identified within this volume, but unless
there is a vast reservoir of cold brown dwarfs invisible to WISE, the final space density of brown dwarfs is
still expected to fall well below that of stars.
We also use these new Y dwarf discoveries, along with newly discovered T dwarfs from WISE, 
to investigate the field substellar mass function. We find that the overall space density of late-T
and early-Y dwarfs matches that from simulations describing the mass function as a power law
with slope $-0.5 < \alpha < 0.0$; however, a power-law may provide a poor
fit to the observed object counts as a function of spectral type because there are tantalizing hints 
that the number of brown dwarfs
continues to rise from late-T to early-Y. More detailed monitoring and characterization of these
Y dwarfs, along with dedicated searches aimed at identifying more examples, are certainly required. 

\end{abstract}

\section{Introduction}

The coldest field brown dwarfs hold important clues that span a variety of astronomical fields. In the field of 
star formation, these coldest brown dwarfs contain a historical record of the formation process at very low 
masses and at epochs many Gyr before the active formation regions we observe today. In the field of planetary 
atmospheric theory, they represent low-temperature atmospheres that can be used as simple test cases for 
predictions because they lack the complications of photochemical processes produced through irradiation by a 
host sun. In the field of exoplanet searches, they provide interesting, nearby targets that may harbor
 planetary systems in their own right.

The Wide-field Infrared Survey Explorer (WISE; \citealt{wright2010}) was built in part to identify these 
coldest brown dwarfs by using their signature methane absorption bands as a photometric diagnostic to 
distinguish them from myriad background sources. Specifically, the shortest wavelength WISE band, hereafter 
denoted W1, has a central wavelength of 3.4 $\mu$m, which falls in the middle of the strong fundamental 
methane absorption band near 3.3 $\mu$m. The second shortest WISE band, hereafter denoted W2, 
has a central wavelength of 4.6 $\mu$m, which detects light arising from deeper, hotter layers in the brown 
dwarf atmosphere; at this wavelength, the atmosphere is fairly transparent to radiation, a direct analog 
being the 5-$\mu$m ``holes'' in the atmosphere of Jupiter (\citealt{westphal1969}). As a result, the W1-W2 
color can be used to identify cold brown dwarfs because that color should be very red. WISE also observes
in two other bandpasses, hereafter denoted as W3 and W4, centered at 12 and 22 $\mu$m, that can be used to 
eliminate sources with longer-wavelength flux inconsistent with that of brown dwarfs. In addition to
color discrimination, WISE also benefits from observations
at wavelengths where cold brown dwarfs are emitting most of their light. 
The all-sky nature of the survey means that the closest and brightest examples in the Solar Neigborhood 
will be imaged.

In this paper we present the discovery of another seven Y dwarfs -- the coldest class of brown dwarfs recognized
(\citealt{cushing2011}) -- which brings the total found with WISE to 
thirteen. In section 2, we discuss their selection as cold 
brown dwarf candidates along with their confirmation as Y dwarfs. In section 3, we discuss further the 
definition of spectral type Y. In 
section 4, we place the Y dwarfs in context with other spectral classes by comparing their relative 
fluxes and space densities, the latter of which is also used to divine clues regarding the shape of the 
low-mass end of the (sub)stellar mass function. In that section, we also compare those results to several 
other studies that have, either directly or indirectly, probed this same mass regime.

\section{New Y Dwarf Discoveries}

\subsection{Candidate Selection}

To support the All-Sky data release in March, 2012, the WISE four-band cryogenic data were reprocessed 
using an improved version of the reduction pipeline, as described in the WISE All-Sky Data Release 
Explanatory Supplement.\footnote{See http://wise2.ipac.caltech.edu/docs/release/allsky/.} The individual 
W1, W2, W3, and W4 frames were rerun and atlas images re-built by stacking the individual, reprocessed 
frames. Source detections were made on the atlas images themselves, and source extractions were made on 
both the individual frames and the atlas images. Our query for cold brown dwarfs used the database of 
extractions from the atlas images, this database being a union of the WISE All-Sky Source Catalog and 
the WISE All-Sky Reject Table. This query attempts to improve upon the search we used earlier 
(\citealt{kirkpatrick2011}) and is aimed toward identifying mid-T or later brown dwarfs for further 
follow-up. The search constraints are given below:

\noindent 1) The W1-W2 color from profile-fit photometry is greater than 2.0 mag, where the W2 measurement 
is an actual detection. (This criterion guarantees that the W2 signal-to-noise ratio is greater than 
three.) As Figure 1 of \cite{kirkpatrick2011} shows, this color is typical of objects of type mid-T 
and later.

\noindent 2) The source is detected with a signal-to-noise ratio greater than three in at least eight 
individual W2 frames going into the coadd stack or, if detected only five, six, or seven times, is still 
detected at a signal-to-noise ratio greater than three in at least 40\% of all frames. This criterion 
is meant to eliminate spurious, co-aligned artifacts in the coadds.

\noindent 3) The source is either undetected in W3 or, if detected, has a W2-W3 color less than 3.5 mag. 
This criterion is meant to eliminate very red extragalactic contaminants or sources embedded in star 
formation regions.

\noindent 4) The source is not flagged as a diffraction spike, star halo, optical ghost, or latent 
artifact in bands W1 and W2. This criterion is meant to remove known spurious sources. Real sources 
that are flagged in W1 and W2 as impacted by spikes, haloes, ghosts, or latents are, however, retained.

\noindent 5) The source is not blended with another source. This criterion is meant to reduce the 
number of objects with poorly determined photometry.

\noindent 6) The source has a reduced ${\chi}^2$ value from profile-fit photometry that lies between 
0.5 and 3.0. This criterion is meant to eliminate sources that are not point-like.

\noindent 7) The source has an absolute Galactic latitude greater than three degrees if its Galactic 
longitude falls within twenty degrees of zero. This criterion is meant to eliminate confused areas 
toward the Galactic Center.

\begin{figure}
\epsscale{0.65}
\figurenum{1}
\plotone{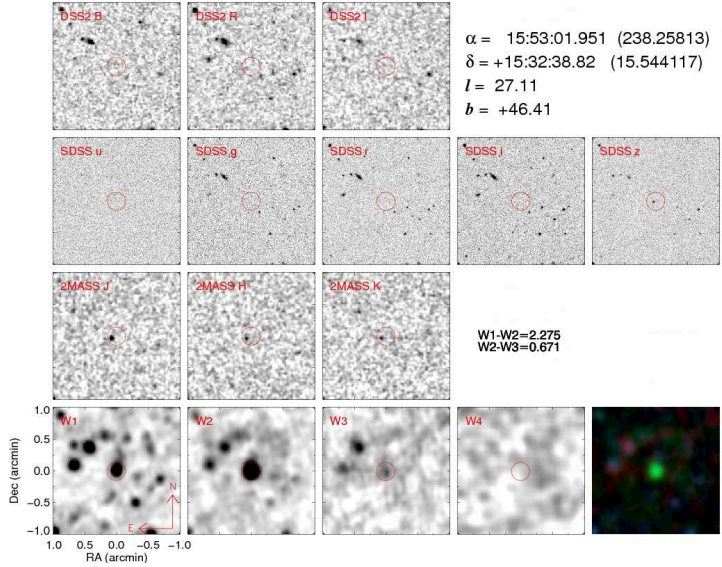}
\caption{Images of one of our brown dwarf candidates, the previously known T7 dwarf 2MASSI J1553022+153236 from \cite{burgasser2002}. Each row in this image represents a different survey and each image shows a 2$\times$2 arcminute region at a different wavelength. The top row shows $B$, $R$, and $I$ images from the Digitized Sky Survey 2 (DSS2). The second row shows $u$, $g$, $r$, $i$, and $z$ images from the Sloan Digital Sky Survey (SDSS; \citealt{york2000}). The third row shows $J$, $H$, and $K_s$ images from the Two-Micron All-Sky Survey (2MASS; \citealt{skrutskie2006}). The last row shows the W1, W2, W3, and W4 images from WISE along with a three-color image comprised of W1 (blue), W2 (green), and W3 (red). Sexagesimal and decimal RA ($\alpha$) and Dec ($\delta$) coordinates of the source are given in the upper right, along with the Galactic longitude ($l$) and latitude ($b$) in decimal degrees. Also shown are the W1-W2 and W2-W3 colors from WISE. A red circle is shown at the location of the candidate identified in WISE. For this object, the source is detected only at SDSS $z$, all three 2MASS bands, and the three shortest wavelength bands of WISE; detections at these wavelengths are common for nearby, bright T dwarfs. Motion is seen from the earliest epochs (2MASS and SDSS) to the latest (WISE), further confirming this as a good candidate. 
\label{FieldCheck}}
\end{figure}

With this list of sources in hand, images of the field were constructed using DSS2 $BRI$ (epoch 
$\sim$1980s), SDSS $ugriz$ (where available; epoch $\sim$2000), 2MASS $JHK_s$ (epoch $\sim$2000), 
and WISE four-band data (epoch $\sim$2010). These image sets, an example of which is shown in 
Figure~\ref{FieldCheck}, show the source across time and across wavelength. By using these images, 
sources that can not possibly be cold brown dwarfs, because they are extended, spurious, or detected 
at wavelengths shortward of 1.0 ${\mu}$m (unless very bright in the WISE W1 and W2 bandpasses), were 
eliminated from further consideration. 

To select objects within 20 pc of the Sun, we further restricted the W2 magnitude 
depending on the color of the source\footnote{Our team also maintains ancillary lists of
candidates with bluer colors or fainter magnitudes, but those are beyond the scope of this paper.}: 
W2 $\le$ 14.5 mag for 2.0 $\le$ W1-W2 $<$ 2.4,
W2 $\le$ 14.8 mag for 2.4 $\le$ W1-W2 $<$ 2.8, and
W2 $\le$ 15.2 mag for 2.8 $\le$ W1-W2 $<$ 2.9.
For colors redder than this -- W1-W2 = 2.9 being the color of the
standard T9 dwarf UGPS J072227.51$-$054031.2 -- no W2 magnitude constraint was
applied because at the time of candidate selection, the absolute magnitudes 
of objects with types $\ge$T9 was very poorly known. This search results in 534 candidates
along with 30 re-discovered T dwarfs identified earlier by other surveys. Candidates have
been placed on our photometry and spectroscopy campaigns, as described in the next section.
To date, 189 of these 534 candidates have been followed up via ground-based or {\it
Hubble Space Telescope} imaging, all 534 are on our {\it Spitzer Space Telescope} follow-up
imaging campaign, and 130 have been observed spectroscopically.
This paper focuses on the 
new Y dwarf discoveries; for newly found T dwarfs, the reader is referred to Mace et al.\ (in prep.), 
Tinney et al.\ (in prep.), and \cite{wright2012}.

\subsection{Follow-up}

\cite{cushing2011} identified the first six Y dwarfs using WISE data, and we confirm seven more here. 
Coordinates and photometry from the WISE All-Sky Release are given in 
Table~\ref{y_dwarf_discoveries_wise} for all thirteen of these Y dwarfs. In order to facilitate 
future investigations, we also provide finder charts for all thirteen in Figure~\ref{Finder0146+4234}. 
These charts show the W1, W2, and W3 discovery images from 
WISE as well as a deep near-infrared (1.2-1.6 $\mu$m) view at higher resolution.  

Photometry of the new Y dwarfs is listed in 
Table~\ref{y_dwarf_discoveries_phot} along with (in some cases, revised) photometry for the six Y 
dwarfs from \cite{cushing2011}. Six ground-based instruments and two space-based facilities -- 
in addition to WISE, whose imaging identified these objects originally -- were used for this 
imaging follow-up. Spectra of the new Y dwarfs were obtained in the $\sim$1.0-1.8 $\mu$m region 
with three different instruments. Details for the spectroscopic observations 
are given in Table~\ref{y_dwarf_discoveries_spec}.

\subsubsection{Mt. Bigelow/2MASS}

The 2MASS camera on the 1.5m Kuiper Telescope on Mt.\ Bigelow, Arizona, observes simultaneously 
in 2MASS $J$, $H$, and $K_s$ filters (\citealt{milligan1996}) using three 256$\times$256 NICMOS3 
arrays. The plate scale for all three arrays is 1$\farcs$65 pixel$^{-1}$, resulting in a 
7\arcmin$\times$7\arcmin\ field of view. The only new Y dwarf in Table~\ref{y_dwarf_discoveries_phot} 
whose photometry we report from the Bigelow/2MASS camera is WISE 0146+4234\footnote{Hereafter, 
we abbreviate the full WISE Jhhmmss.ss$\pm$ddmmss.s designations in the text as WISE hhmm$\pm$ddmm.}. 
Data acquisition and reduction for this instrument have been described earlier in \cite{kirkpatrick2011}.

\subsubsection{AAT/IRIS2}

The IRIS2 instrument (\citealt{tinney2004}) at the 
3.9m Anglo-Australian Telescope (AAT) at Siding Spring Observatory, Australia, provides wide-field 
imaging (7$\farcm$7$\times$7$\farcm$7) using a 1024$\times$1024 (0$\farcs$4486 pixel$^{-1}$) 
Rockwell HAWAII-1 HgCdTe infrared detector. Our observation of WISE 2220$-$3628 used only 
the $J$ filter, which is on the MKO-NIR system (\citealt{tokunaga2002}). Data collection and 
reduction for this instrument are described in Tinney et al.\ (in prep.).

\subsubsection{CTIO/NEWFIRM}

The NOAO Extremely Wide Field Infrared Imager (NEWFIRM; \citealt{swaters2009}) at the 4m Victor M.\ 
Blanco Telescope on Cerro Tololo, Chile, uses four 2048$\times$2048 InSb arrays arranged in a 
2$\times$2 grid.  With a pixel scale of 0$\farcs$40 pixel$^{-1}$, this grid covers a total field 
of view of 27$\farcm$6$\times$27$\farcm$6. Only one of our new Y dwarfs, WISE 0734$-$7157, was 
acquired with this instrument and it was observed only at $J$ band, which is on the MKO-NIR system. 
Observing and reduction strategies are described in \cite{kirkpatrick2011}.

\subsubsection{SOAR/SpartanIRC}

The Spartan Infrared Camera (SpartanIRC; \citealt{loh2004}) at the 4.1m Southern Astrophysics 
Research (SOAR) Telescope on Cerro Pach{\'o}n, Chile, uses four 2048$\times$2048-pixel HAWAII-2 
arrays arranged in a 2$\times$2 grid. The field of view can be set to cover either a 
3\arcmin$\times$3\arcmin\ (0$\farcs$043 pixel$^{-1}$) or 5\arcmin$\times$5\arcmin\ (0$\farcs$073 
pixel$^{-1}$) area per array. Our only observation, of WISE 0713$-$2917, was done with the larger 
field of view in the $J$ and $H$ filters, which are on the MKO-NIR 
system. Observing strategy and data reductions followed the same prescription discussed in 
\cite{burgasser2011}.

\subsubsection{SOAR/OSIRIS}

The Ohio State Infrared Imager/Spectrometer (OSIRIS), also at SOAR, uses a 1024$\times$1024 HAWAII HgCdTe array. The 
field of view can be set to cover either a 2$\farcm$4$\times$2$\farcm$4 (0$\farcs$139 pixel$^{-1}$) 
or 5$\farcm$6$\times$5$\farcm$6 (0$\farcs$331 pixel$^{-1}$) area. The Y dwarf WISE 0713$-$2917 was
observed at both $J$ and $H$, and the Y dwarf WISE 2220$-$3628 was observed only at $H$ band. These are Barr 
filters; for $H$ band, the filter curve\footnote{See http://www.ctio.noao.edu/instruments/ir\_instruments/osiris2soar/config/index.html.} 
has half-power points near 1.48 and 1.78 $\mu$m. Comparison of this filter curve to the filter 
curves illustrated in figure 4 of \cite{bessell2005} shows that it is very similar to the 
MKO-NIR $H$-band filter. Observing strategy and data reductions followed the same prescription as those
of SpartanIRC (discussed in \citealt{burgasser2011}).

\subsubsection{Magellan/PANIC}

Persson's Auxiliary Nasmyth Infrared Camera (PANIC; \citealt{martini2004}) at the 6.5m Magellan 
Baade Telescope on Cerro Manqui at the Las Campanas Observatory, Chile, has a 1024$\times$1024 
HAWAII array with a plate scale of 0$\farcs$125 pixel$^{-1}$, resulting in a 2\arcmin$\times$2\arcmin\ 
field of view. Observations of three of our new Y dwarfs -- WISE 0350$-$5658, WISE 0359$-$5401, 
and WISE 0535$-$7500 -- were obtained at $J$ and $H$ bands on the Carnegie (essentially MKO-NIR) 
filter system. Details regarding standard data acquisition and reduction methods can be found in 
\cite{kirkpatrick2011}.

\subsubsection{Spitzer/IRAC}

The Infrared Array Camera (IRAC; \citealt{fazio2004}) onboard the {\it Spitzer} Space Telescope 
employs 256$\times$256-pixel detector arrays to image a field of view of 5$\farcm$2$\times$5$\farcm$2 
(1$\farcs$2 pixel$^{-1}$). IRAC was used during the warm {\it Spitzer} mission to obtain deeper 
photometry in its 3.6 and 4.5 $\mu$m channels (hereafter, ch1 and ch2, respectively) than WISE was 
able to take in its W1 and W2 bands. Such observations give us more definitive colors over this 
wavelength regime because the WISE W1-W2 colors (Table~\ref{y_dwarf_discoveries_wise}) are limits 
only. These observations were made as part of Cycle 7 and Cycle 8 programs 70062 and 80109 
(Kirkpatrick, PI) and include all thirteen of our Y dwarf discoveries (Table~\ref{y_dwarf_discoveries_phot}). 
Our standard data acquisition and reduction methodology for IRAC observations is outlined in 
\cite{kirkpatrick2011}.

\subsubsection{HST/WFC3}

The Wide Field Camera 3 (WFC3\footnote{See http://www.stsci.edu/hst/wfc3.}) onboard the {\it Hubble} 
Space Telescope employs a 1024$\times$1024 HgCdTe detector with a plate scale of 0$\farcs$13 pixel$^{-1}$ 
to image a field of view of 123\arcsec$\times$126\arcsec. It was used to obtain deep near-infrared 
photometry in the F140W filter (a broad bandpass encompassing most of the $J$ and $H$ bands) for 
seven of our Y dwarfs, as listed in Table~\ref{y_dwarf_discoveries_phot}. These observations were 
made as part of our Cycle 18 program 12330 (Kirkpatrick, PI). Photometry was measured on the drizzled
images, and magnitudes were measured on the Vega system. For the brighter sources, a comparison of the photometry measured from
the drizzled images to that measured from the individual, direct images showed a systematic difference
that was sometimes as large as 0.20 mag. To be conservative, we have adopted 0.20 mag as the uncertainty 
for all sources since this systematic difference is believed to be the dominant error term.

This same program also used the G141 grism to acquire slitless spectroscopy over the 1.1-1.7 $\mu$m 
range. This was done to confirm three new brown dwarf discoveries as well as to obtain better signal 
for a fourth, WISE 2056+1459, in addition to those WFC3 spectra already discussed in \cite{cushing2011} 
and \cite{kirkpatrick2011}. Each target was observed over a single orbit, and the integration times 
were selected to best fill the time available between Earth occultations. Integration times were 
2212s for all objects except WISE 2056+1459, for which only 2012s could be obtained. Details of 
data acquisition and reduction are given in \cite{kirkpatrick2011}.
Spectra of the program objects were free of contamination by other field spectra and had uncomplicated 
reductions except for WISE 0535$-$7500, whose first-order spectrum falls coincident with the 
second-order spectrum of a field star. The blended spectrum of this source was extracted in the
same way as the others, and a correction (discussed in Section 3) was applied to mitigate the 
effects of the contaminating object.

\subsubsection{Keck/NIRSPEC}

The Near-Infrared Spectrometer (NIRSPEC, \citealt{mclean1998,mclean2000}) at the 10m W.~M.~Keck 
Observatory on Mauna Kea, Hawai'i, was used to confirm three of our new Y dwarfs and to obtain 
additional signal for a fourth, WISE 2056+1459 (\citealt{cushing2011}). For spectroscopy, NIRSPEC 
uses a 1024$\times$1024 InSb array. In low-resolution mode, use of the 42\arcsec$\times$0${\farcs}$38 
slit results in a resolving power of R~$\equiv~{\lambda}/{\Delta}{\lambda}~{\approx}~2500$. Our 
brown dwarf candidates were observed in either or both of the N3 and N5 configurations (see 
\citealt{mclean2003}) that cover part of the $J$-band window from 1.15 to 1.35 $\mu$m and part 
of the $H$-band window from 1.5 to 1.8 $\mu$m. Standard data acquisition and reduction techniques, 
as described in Mace et al.\ (in prep.), were used.

\subsubsection{Magellan/FIRE}

The Folded-port Infrared Echellette (FIRE; \citealt{simcoe2008}, \citealt{simcoe2010}) at the 6.5m 
Walter Baade Telescope on Cerro Manqui at the Las Campanas Observatory, Chile, uses a 2048$\times$2048 
HAWAII-2RG array. In prism mode, it covers a wavelength range from 0.8 to 2.5 $\mu$m at a resolution 
ranging from R=500 at $J$-band to R=300 at $K$-band for a slit width of 0$\farcs$6. FIRE was used to 
confirm WISE 0734$-$7157 as a Y dwarf\footnote{Because of the faintness of this source at $K$ band, 
the FIRE spectrum beyond 1.65 $\mu$m is not plotted in the figures as it is very noisy.}. For 
standard data acquisition and reduction techniques for FIRE, refer to the discussion in 
\cite{kirkpatrick2011}.

\section{The Classification of Y Dwarfs}

\subsection{Classifying the New Discoveries}

Our new spectra, along with those from \cite{cushing2011}, are plotted in Figures~\ref{YDwarfs1}, 
\ref{YDwarfs2} and \ref{YDwarfs3}. In Figures~\ref{YDwarfs1} and \ref{YDwarfs2} we show all of the 
Y dwarfs except WISE 0535$-$7500 (which is plotted separately in Figure~\ref{YDwarfs3}) and compare 
them to the T9.5 dwarf WISE 0148$-$7202 and other Y dwarfs. In Figure~\ref{YDwarfs3} we plot two 
versions of the spectrum of WISE 0535$-$7500. In the first, we show the contaminated spectrum as 
extracted by the {\it aXe} software from the {\it Hubble Space Telescope}. 
In the second, we plot our attempt at correcting the spectrum 
for the light of the contaminating object. This simplistic correction involves subtracting off a 
linear slope so that the mean level in the water absorption bands is zero. The resulting, quasi-corrected 
spectrum is sufficient to provide a crude classification despite the problems with data acquisition.

After examining these plots, several facts become apparent: (1) All seven of the new discoveries have 
$J$-band ($\sim$1.27 $\mu$m) flux peaks as narrow or narrower than the Y0 standard, WISE 1738+2732, proposed by 
\cite{cushing2011}. The T9.5 dwarf, in contrast, has a wider $J$-band peak than any of these. This 
confirms that all are Y dwarfs, as defined by \cite{cushing2011}. (2) The spectra of WISE 1541$-$2250 and WISE 0350$-$5658 have 
distinctly narrower $J$-band peaks than the Y0 standard itself, meaning that both of these objects 
should be typed later than Y0. The plot shown in Figure~\ref{YDwarfs_Jcomparison} suggests that the 
$J$-band peak of WISE 0350$-$5658 is sufficiently narrower than WISE 1738+2732 that it should be 
typed a full subclass later. We therefore propose WISE 0350$-$5658 to be the tentative Y1 spectral 
standard, despite its southerly declination, until other Y1 dwarfs are identified. This also means 
that WISE 1541$-$2250, which has a slightly broader $J$-band peak, should be re-typed from Y0 to Y0.5. (3) 
The spectrum of WISE 1828+2650, though having a low signal-to-noise ratio, is still in a class by 
itself due to the near-equal heights of the $J$- and $H$-band ($\sim$1.58 $\mu$m) peaks. Because it is so different from 
all of the other spectra, we re-classify it to be $\ge$Y2.

\subsection{Revisiting the Establishment of the Y Dwarf Class}

Although \cite{cushing2011} discussed in detail the establishment of the new Y dwarf spectral class,
it is worth revisiting this based on the latest discoveries. We compare our set
of observational Y dwarf spectra to theoretical 
expectations prior to the launch of WISE. Specifically, \cite{burrows2003} used atmospheric
models covering this temperature regime to cite five possible triggers that
could lead to the introduction of a new class beyond the T dwarfs. We investigate each of these 
possibilities below:

\noindent (1) The disappearance of alkali resonance lines near 450K. The resonance lines of \ion{Na}{1}
and \ion{K}{1} fall at optical wavelengths, where cold brown dwarfs have little flux, but the
broad wings of these lines extend very far from the line cores and are believed to have an
influence on the emergent flux as longward as 1 $\mu$m (\citealt{burrows-volobuyev2003}).
As discussed in \cite{cushing2011}, the $Y$-band ($\sim$1.07 $\mu$m) peaks for the Y dwarfs whose spectra cover that
region generally appear to be as high or higher than the $J$-band peaks in units of $F_\lambda$.
This effect is not seen in the spectra of late-T dwarfs (see, e.g., \citealt{kirkpatrick2011}). 
A harbinger of this effect -- seen as a blueward trend of $Y$-$J$ color in the mid- to late-T dwarf regime --
was noted by \cite{leggett2010} 
and \cite{burningham2010} and ascribed to a brightening of the $Y$-band peak due to reduced 
absorption by the wings of the \ion{K}{1} as atomic potassium begins to form into KCl at
cooler temperatures (\citealt{lodders1999}). 
Although indirect, this is evidence that we have
pushed into the regime where these alkali lines have lost prominence.
Further spectroscopic investigation  
at higher signal-to-noise levels is possible at $Y$-band using the G102 grism onboard ${\it HST}$/WFC3.

\noindent (2) Water cloud formation below 400-500K. Although long-predicted to be a possible trigger
of a new spectral class at low temperatures, \cite{burrows2003} found that the formation of these
clouds has very little effect on the emergent spectra in their models. Marley (priv.\ comm.) also
finds little effect shortward of 2 $\mu$m but finds that at longer wavelengths, for objects with
$\la$300K, the effects are quite pronounced. The importance of these clouds should be re-investigated
once newer atmospheric models are published for the coldest brown dwarfs.

\noindent (3) The emergence of ammonia absorption below 2.5 $\mu$m. As shown in figure 5 of
\cite{kirkpatrick2008-coolstars}, the theoretical spectra of \cite{burrows2003} indicate that
absorption bands of NH$_3$, which first appear in the mid-infrared (10.5 $\mu$m) near the L/T
transition (\citealt{cushing2006}),
finally appear at the $H$ and $K$ bands starting near 800K and become
prominent by 450K. As shown in figure 5 of \cite{cushing2011}, the $H$-band ammonia feature --
the easier of these two to detect because the $K$-band feature falls within the telluric water
band -- is itself confused with overlying absorption by H$_2$O and CH$_4$ in the brown dwarf 
atmosphere. Nonetheless, signs of a possible
NH$_3$ signature were noted by \cite{cushing2011} in the 1.53-1.58 $\mu$m region of WISE 1738+2732.
This region falls in an area of extremely low flux, so exquisite signal-to-noise
is needed to detect it.  Unfortunately, none of the spectra 
of our new objects has sufficient signal to investigate this feature further. Analyzing the
near-infrared spectra of these objects for the presence of NH$_3$ bands will likely require
higher resolution spectra than can be presently obtained. 

\noindent (4) Collapse below 350K of the optical and near-infrared fluxes, relative to those $\ga$5 $\mu$m.
\cite{burrows2003} state that this collapse in flux would manifest itself as a 
reversal of the blueward trend of $J-K$ (or $J-H$) colors.
Figure 11 in \cite{cushing2011} shows the $J-H$ color 
(on the MKO-NIR system) as a function of spectral type. That figure shows that the $J-H$ color stagnates 
near $-$0.4 mag for late-T dwarfs then appears to turn to the red starting at Y0, although there 
are some blue outliers at Y0. With our new Y dwarf discoveries, 
new T dwarf discoveries from Mace (et al.),
additional 
near-infrared photometry, and revised photometric reductions of previous Y dwarf discoveries (see 
footnotes to Table~\ref{y_dwarf_discoveries_phot}), we can revisit this plot. This new photometry, 
shown in Figure~\ref{YDwarfs_Colors2}, more clearly shows the reversal of the $J-H$ color. After 
trending to the blue from early- to mid-T and stagnating for late-T types, the $J-H$ color turns 
to the red starting near Y0. Later Y dwarfs, such as the Y0.5 dwarf 
WISE 1541$-$2250 ($J-H >$ 0.54 mag) and the $\ge$Y2 dwarf WISE 1828+2650 ($J-H$ = 0.72$\pm$0.42 
mag) are roughly one magnitude redder than late-T dwarfs.

\noindent (5) The shift in position of the $\sim$5 $\mu$m peak.
As shown in figures 7 and 8 of \cite{kirkpatrick2011}, the $J-$W2 and $H-$W2 colors continue to 
increase from mid-T to early-Y. Figure 11 from that paper shows an indication that the {\it Spitzer} 
ch1$-$ch2 color may reverse in the Y sequence, but this was based solely on the ch1$-$ch2 color of 
WISE 1828+2650. We can also revisit this trend using newly discovered Y dwarfs. Figure~\ref{YDwarfs_Colors} 
shows the $J-$W2, F140W$-$W2 and $H-$W2 colors as a function of the ch1$-$ch2 color\footnote{Plots showing
the $J-$ch2, F140W$-$ch2, and $H-$ch2 colors would look nearly identical. The W2 magnitudes listed in
Table~\ref{y_dwarf_discoveries_wise} are nearly identical to the ch2 magnitudes in Table~\ref{y_dwarf_discoveries_phot}, though the latter have errors
generally two to three times smaller. However, the error in the color measurements is dominated by
the $J$, F140W, or $H$ terms anyway.}. 
The ch1$-$ch2 color
shows a broad range (of up to a magnitude) for Y0 dwarfs; similar scatter is seen in the ch1$-$ch2
colors of late T dwarfs, which \cite{leggett2010} attribute to gravity and/or metallicity effects.
Dwarfs classified as Y1 or later
are generally redder than these. It should be noted that the $\ge$Y2 dwarf, WISE 1828+2650, indicates a 
turn to the blue in ch1$-$ch2 color at later types. (The $J-$W2, F140W$-$W2, and $H-$W2 colors, nevertheless, still 
tend to run redder with advancing spectral type and may serve as a proxy for temperature for these early-Y dwarfs.)

WISE 1828+2650 satisfies the last two of the possible trigger conditions discussed above.
The stark contrast between its spectral morphology and that of T dwarfs -- namely, the near-equal heights of the
$J$ and $H$ peaks that is manifested as a turn to the red in the $J-H$ color -- satisfies the criteria needed to 
define a new spectral class. Moreover, Beichman et al.\ have recently measured the
trigonometric parallax for this object, establishing it as an instrinsically dim source (see
section 4.3), and model fits by \cite{cushing2011} suggest that its effective temperature is below 300K.
With more evidence now in hand, we still reach the same conclusion as \cite{cushing2011} that WISE 1828+2650
should be considered as the archetypal Y dwarf.

Given that the ammonia absorption bands in the near-infrared are obscured by other strong absorption bands,
we lack a clear signature that defines the exact onset of the Y class. A gradual change in spectral 
morphology is the norm, however, at the boundary between spectral classes. For example, there is very little change in the
optical morphology of a late-K dwarf and an early-M dwarf, even though mid-K and mid-M spectra are 
markedly different, and the same is true at the boundary between M and L classes. The L/T transition
is unusual in its sudden appearance of a major absorption species that radically alters the appearance
of the spectra (over what we now know is a very small temperature range; see figure 8 of \citealt{kirkpatrick2005}). 
For Y dwarfs, we see a gradual
change in the spectral morphology from late-T to early-Y, even though the spectra of a mid-T dwarf and that 
of WISE 1828+2650 are very different. We have further evidence that this gradual
change continues down the Y sequence, as well: the spectrum of our new Y1 dwarf WISE 0350$-$5658 has as an $H$-band peak relative to that
of the $J$-band peak that is higher than in the Y0 standard and this presages the effect of equal-peak
heights seen in WISE 1828+2650. We therefore see no reason to deviate from the Y dwarf classification
scheme presented in \cite{cushing2011}, which has so far been robust to new discoveries.

\subsection{A New Spectral Index for Early-Y Dwarfs?}

For early-Y dwarfs, we have based our classifications on the narrowness of the $J$-band peak using
by-eye comparisons to our Y0 and Y1 spectral standards. Can a spectral index be created that
distills this same information? A pre-existing index, dubbed W$_J$, was developed by 
\cite{warren2007} to measure this narrowness of $J$-band for late-T dwarfs, but we find that
it has problems in the Y dwarf regime. (See figure 7 of \citealt{cushing2011} for a graphical representation of this index
along with other indices measured for T dwarfs.)
All of the Y dwarfs have a W$_J$ index of $\sim$0.1,
except for WISE 1828+2650, which is very noisy. Because the $J$-band peak 
is so narrow for Y dwarfs, the numerator of the W$_J$ index (which is the flux integrated from 1.18 to 1.23 
$\mu$m) is nearing zero. 
A better measure of the $J$-band peak can be obtained by moving this 
region of integration longward so that it still falls in a region with measurable flux nearer the peak.
Also, the region used in the denominator of the index (1.26 to 1.285 $\mu$m), which measures the flux in 
the peak itself, as well as the region used in the numerator (to measure the blueward wing) can be narrowed to 
reflect the fact that the opacity hole in the spectrum has narrowed for the Y dwarfs. 

Mace et al.\ define
a new index, called $J$-narrow, which is the ratio of the median flux over the 1.245-1.260 $\mu$m
region to the median flux over 1.260-1.275 $\mu$m. We show the results of measuring this
index on our collection of WISE-discovered T and Y dwarfs in Figure~\ref{YDwarfs_Indices}. Note that 
for most of the T dwarfs, $J$-narrow has a value greater than 0.85. Y dwarfs have values less than this,
the Y0 standard WISE 1728+2732 being the least narrow Y dwarf as measured by this index. (This is because
our by-eye classification requires an object to have a $J$-band peak as least as narrow as that of the 
Y0 standard for it to be classified as a Y dwarf.)
Whittling the wavelength region over which the index operates, though necessary, nonetheless comes at a price: 
high signal-to-noise spectra are required to make the index measurements robust. This is a difficult requirement 
to meet because Y dwarfs are notoriously hard to observe given their intrinsic faintness and the ones plotted
in Figure~\ref{YDwarfs_Indices} are among the closest, brightest Y dwarfs over the entire sky. Even though we
present this new index for completeness, 
{\it we strongly recommend that researchers classify their spectra by overplotting the late-T and 
early-Y standards rather than resorting to the use of spectral indices, since the latter technique 
is far more prone to error when spectra have low signal-to-noise.}

\section{Y Dwarfs in the Larger Context}

As theorized, old brown dwarfs in the field population were expected to be very faint given the fact that they have no 
sustained source of energy in their interiors. The faintness of these objects as a function of 
time could be modeled based on interior physics (e.g., figure 11 of \citealt{burrows1997}), but the more difficult issue was predicting how 
frequently they might occur in nature, if at all. Using the Y dwarfs discovered by WISE and 
monitoring their astrometry to measure trigonometric parallaxes, we are able to demonstrate 
just how faint these objects are and can begin to put more solid limits on the Y dwarf space 
density and the shape of the mass function in the low-temperature, low-mass realm.

\subsection{Absolute Magnitudes and the H-R Diagram}

To understand the intrinsic faintness of Y dwarfs, we have constructed a Hertzsprung-Russell (H-R) diagram
at $H$-band that contains a selection of main sequence stars and brown dwarfs with spectral types running the gamut from O 
through Y. Objects were required to have high quality parallaxes and be free of complications such as 
being heavily reddened (which is mitigated by chosing only the most nearby examples)
or belonging to a close multiple system. Our aim was to choose, if possible, at least twenty-five 
objects in each spectral class, spread evenly so that all integral subclasses were covered. All $H$-band 
magnitudes and errors were taken from 2MASS. Additional details are given below:

\noindent {\it O, B, and A dwarfs}: Using Jim Kaler's ``Stars'' website\footnote{See http://stars.astro.illinois.edu/sow/sowlist.html.},
we selected objects identified as O dwarfs and retained those having trigonometric parallaxes from Hipparcos
(\citealt{vanleeuwen2007}) that place them within 600 pc of the Sun and parallax errors less than one 
milliarcsecond (mas). Only ten O dwarfs survived the cut, the earliest being the O7 V star 15 Monocerotis.
Using the same website and Hipparcos parallaxes, we selected B dwarfs lying within 500 pc and A dwarfs 
lying within 100 pc. Retaining only those with parallax errors under $\sim$1 mas resulted in a sample of
twenty-six B stars and forty-two A stars.

\noindent {\it F, G, and K dwarfs}: Using SIMBAD, we selected F, G, and K stars whose Hipparcos parallaxes
(\citealt{vanleeuwen2007})
place them within 20.0, 18.2, and 13.4 pc of the Sun (i.e., parallax values exceeding 50, 55, and 75 mas), 
respectively. We retained those that Jim Kaler's ``Stars'' website confirms as having dwarf
lumninosity classes. Further requiring that the Hipparcos parallaxes errors are less than $\sim$0.5 mas
resulted in a list of thirty-five F dwarfs, fifty-three G dwarfs, and twenty-eight K dwarfs. 

\noindent {\it M dwarfs}: We selected M dwarfs from the Research Constortium On Nearby Stars' (RECONS) web page that
lists the hundred nearest stellar systems to the Sun\footnote{The list was consulted in late December, 2011. 
See http://www.recons.org/.}. Dropping objects where the 2MASS $H$-band photometry or measured spectral type
is a composite of multiple components resulted in a list of seventy-two M dwarfs.

\noindent {\it L and T dwarfs}: We selected from the literature those L and T dwarfs having measured trigonometric parallaxes.
The list is available as Table 5 of \cite{kirkpatrick2011}. Retaining only those not known to be multiple 
systems and having well measured 2MASS $H$-band magnitudes results in twenty L dwarfs and twenty-one T dwarfs.

\noindent {\it Y dwarfs}: Due to the paucity of parallax data currently available for Y dwarfs, we retained all
Y dwarfs from Marsh et al.\ (in prep.) and Beichman et al.\ (in prep.) having trigonometric parallax measurements
at least five times the error. This selection yields four Y dwarfs.

The compiled list is illustrated in Figure~\ref{MH_type}. The trend of ever-dimming $H$-band magnitude as a 
function of later spectral type shows two well known inflection points. The first of these occurs 
at early-M and is a result of hydrogen associating into H$_2$ at temperatures below $\sim$4000K 
(\citealt{mould1976}; \citealt{mould-hyland1976}). The other 
inflection point, at late-L to mid-T, is a flattening or brightening of the $H$-band flux over 
a range of spectral types near the L/T transition (see, e.g., \citealt{looper2008},
\citealt{dahn2002}, \citealt{vrba2004}, \citealt{knapp2004}; the brightening is 
even more dramatic in the $J$-band). The physical cause for this brightening may be due to patchy clouds
in the atmospheres of these objects (\citealt{marley2010}).
Below this second inflection point, the $H$-band magnitude plummets. By early-Y the absolute $H$-band 
magnitudes are $\sim$30 mags (or 12 orders of magnitude in flux) fainter than those of late-O dwarfs. 
As discussed later in this paper, the trend of absolute $H$-band flux may show another inflection
point at early-Y, although more data are required to check this further.

\subsection{Y Dwarf Number Density and the 8-pc Census}

To understand the importance of Y dwarfs in the Milky Way, we consider 
an all-sky, volume-limited sample of the Solar Neighborhood with
which to compare the frequency of Y dwarfs relative to other spectral types. 
Previous authors have considered different volumes as defining the immediate Solar Neighborhood.  
Building on earlier work\footnote{\cite{hertzsprung1907} 
was the genesis of the H-R diagram; see \cite{batten1998} for
more on the historical importance of this paper.} by \cite{hertzsprung1907,hertzsprung1922}, 
\cite{vandekamp1930,vandekamp1940,vandekamp1945,vandekamp1953,vandekamp1969,vandekamp1971} 
considered the sample out to $\sim$5 pc; 
\cite{kuiper1942} considered a volume out $\sim$10 pc; and 
\cite{gliese1956,gliese1969} and \cite{gliese1979} considered a distance limit of $\sim$20 pc, although this
was extended to 25 pc in \cite{gliese1991}. For our purposes, we will consider a distance limit of 8 pc
because this bounds a volume with a sufficient number of objects ($\sim$250) to provide adequate statistics
across spectral classes. 

Table~\ref{8pc_census} gives
our update of all stars and brown dwarfs known or suspected to lie within 8 pc. This list relies heavily on 
previous work, most notably the list compiled by \cite{reid1997} with updates in \cite{reid2004} and
Reid (priv.\ comm.) and the list of the hundred nearest stars compiled by RECONS
at their website. The papers by Reid et al.\ consider only objects with Dec $> -30^o$
and do not include several recently discovered, low-luminosity objects uncovered by surveys such as 2MASS, SDSS, DENIS, UKIDSS, and WISE;
the RECONS list includes only objects with precisely determined trigonometric parallaxes and only objects within roughly 6.7 pc 
of the Sun. Therefore, we have combed the literature to uncover more newly discovered objects, suspects with unknown or poorly
measured parallaxes, or objects near the outer limits of the 8-pc volume.
Other updates in Table~\ref{8pc_census} include better characterization of previously known or newly identified multiple
systems and consistent, MK-based spectral types for stars earlier than late-K (see \citealt{gray2003} and
\citealt{gray2006}). Further details on each column are given below.

Column 1 of Table~\ref{8pc_census} is intended to provide homage to the original discoverer or survey/mission
responsible for first identifying the star as a nearby object. Exceptions are made in the case of stars
with common names (e.g., Altair, Fomalhaut, and Vega), Bayer designations (e.g., $\alpha$ Cen A and B, p Eri AB),
Flamsteed designations (e.g., 36 Oph ABC), or designations in old stellar catalogs
(e.g., Lalande 21185, Lacaille 9352, AC+79 3888). In some cases, it is difficult to determine whether Willem Luyten or
the Lowell Observatory group led by Henry Giclas was the first to discover an object because both groups were
undergoing photographic proper motion surveys simultaneously. For such objects, the Luyten designation is 
used if the Lowell Observatory group lists one in its cross-references (\citealt{giclas1971,giclas1978}),
and the Giclas number is used if no Luyten designation is given.\footnote{Luyten did not usually provide cross-references 
to the Giclas numbers. The Lowell group aimed to acknowledge Luyten, however, when they rediscovered one of his
proper motion objects. For more details, see the interview of Dr.\ Henry Giclas by Robert Smith
on August 12, 1987, Niels Bohr Library \& Archives, American Institute of Physics, College Park, MD, USA,
http://www.aip.org/history/ohilist/5022.html.} Column 2 lists alternate names.

Column 3 of Table~\ref{8pc_census} gives the running number in \cite{gliese1956,gliese1969} or \cite{gliese1979}.
SIMBAD has made it common practice to identify objects from any of these papers with a prefix of ``GJ'', but this
was originally meant for objects only from \cite{gliese1979}. Moreover, for new objects (identified only by ``NN'') in 
\cite{gliese1991}, SIMBAD has created its own numbering scheme; 
identifiers with GJ numbers higher than GJ 2159 are solely a SIMBAD creation\footnote{Likewise, we do not use
LHS numbers (\citealt{luyten1979}) higher than LHS 5413 because these are also a SIMBAD creation.}. 
\cite{gliese1991} do not provide catalog numbers for new
objects nor are ones needed because all have published, well recognized names. 
Such was not necessarily the case for earlier versions of the catalog -- researchers in the early 1900's, notably Robert Innes, did not 
always affix numbers or names to their discoveries, and readers may not have had, as we do today, 
easy access to discovery papers.
In deference to the intent of the original publications, we give a prefix of ``Gl'' to objects from 
\cite{gliese1956,gliese1969} and ``GJ'' only to those objects
in \cite{gliese1979}.  For other objects, the field in column 3 is left blank.

Columns 4-6 of Table~\ref{8pc_census} give the measured trigonometric parallax, its error, and the reference
for the measurement. If more than one group has measured the parallax, we list the published measurement with 
the smallest quoted error. For cases where the error in column 5 is blank, the value in column 4 is actually a 
spectrophotometric estimate based on the magnitude of the object and its spectral type. Columns 7-8 give
the spectral type and its reference; cases where the spectral type is estimated are annotated as such.
Column 9 provides an abbreviated sexagesimal J2000 position for the object in the form hhmm$\pm$ddss.
Column 10 gives the rank of the system in distance from the Sun, and column 11 gives the individual rank of
each object; for example, Sirius A is in the fifth closest stellar system to the Sun but is tied with
its companion, Sirius B, as the seventh closest star.

For reference, objects included in previous compilations of the 8-pc sample but now believed to lie beyond 8 pc
are listed in Table~\ref{8pc_census_out}. Among these is the intriguing multiple system $\xi$ UMa (Alula Australis), in
which \cite{wright2012} have announced the discovery by WISE of a widely separated T8.5 companion.

Figure~\ref{8pc_sample} graphically illustrates the 8-pc sample of Table~\ref{8pc_census} and shows that the Solar Neighborhood is
dominated by M dwarfs. There are twice as many M dwarfs known as there are all other spectral types combined,
although they comprise slightly less than half of the total stellar mass. Dwarfs of types L, T, and Y are much less
common. L dwarfs are especially rare -- only three examples are known within 8 pc -- and represent a mix of old
stars at the low-mass end of the stellar mass function and old brown dwarfs at the high-mass end of the substellar
mass function\footnote{Using the \cite{burgasser2004,burgasser2007} luminosity function simulations,
we find that an $\alpha = 0$ power law would suggest approximately 12 L dwarfs within
this 8-pc volume. It is difficult to imagine that previous surveys have missed three quarters of all the nearby L dwarfs,
so we conclude either that the appropriate value of $\alpha$ is less than zero and/or that a power-law is a poor 
representation for the mass function of brown dwarfs. We return to this point in the next section}. 
The numbers of brown dwarfs then rises at later types -- 22 known T dwarfs and 8 known Y dwarfs are 
thought to lie within this 8-pc volume.  

The statistics for T and Y dwarfs are still incomplete, however. For T dwarfs, a few late-type isolated objects may 
yet be found, and it is expected that T dwarf companions to some of the higher mass stars will continue to be 
uncovered. The number of Y dwarfs is incomplete because follow-up of Y dwarf candidates from WISE is still in
its infancy and because the coldest Y dwarfs are likely beyond the detection theshold of all existing surveys.
As discussed in the next section, distances to the Y dwarfs are only now being measured for the first time, so
many of the Y dwarf distances in Table~\ref{8pc_census} are estimates only. 

\subsection{The Low-mass End of the Field Brown Dwarf Mass Function}

We now concentrate on the low-mass tail of the field brown dwarf distribution in an updated attempt to determine the shape
of the substellar mass function and its low-mass cutoff. As was done in \cite{kirkpatrick2011}, we focus on objects 
with spectral types of T6 or later. To sample a sufficiently large number of objects at these types,
we consider distances larger than the 8-pc limit of the previous section, but we must keep in mind that samples should
be reasonably complete (or completable) out to the distance limit we choose. Toward this end, we will use the depth of the
WISE all-sky survey as the arbiter of the distance to which a complete census can be obtained, since it is the 
survey most capable of finding these coldest brown dwarfs.
In \cite{kirkpatrick2011} we found that a distance limit of 20 pc worked well for integral T6-T6.5, T7-T7.5, and T8-T8.5 bins. 
At T9-T9.5, it
was shown that WISE may only fully sample the entire sky out to $\sim$15 pc, so this limit was chosen for that
spectral bin. The limit for Y dwarfs was chosen to be 10 pc, which we retain here and discuss in more detail in the discussion 
that follows. 

Many of the objects will not have trigonometric parallax measurements, so distances must be estimated
spectrophotometrically -- i.e., each object's apparent magnitude will be compared to the absolute
magnitudes computed for similarly typed objects whose distances have been measured. Apparent magnitudes for these cold brown
dwarfs are most easily measured from the ground in the $J$ and $H$ bands and from WISE at W2. Unfortunately,
due to big differences in the $J$-band filters chosen for the two most popular filter systems (2MASS and the
MKO-NIR system), fluxes can vary by several tenths of a magnitude for the same object (\citealt{stephens2004})
because the spectral energy distributions of cold brown dwarfs are so complex at these same wavelengths. Fortunately,
the $H$-band filters are very similar and as a result, $H$-band measurements show little variation between
systems (\citealt{stephens2004}). We will therefore use $H$ and W2 photometry to estimate distances to
objects lacking parallax information.

Table~\ref{TYparallaxes} presents a compilation of dwarfs that have measured trigonometric parallaxes,
photometry at $H$ and/or W2, and types later than T4. Trigonometric parallaxes for the Y dwarfs are
taken from Marsh et al.\ (in prep.) and Beichman et al.\ (in prep.), although three objects from Marsh et al.\ 
-- WISE 0359$-$5401, WISE 1541$-$2250, and WISE 2056+1459 -- are omitted here because their parallax
values are roughly the size of the errors themselves. Unlike the other objects in these
papers, these three objects lack measurements near the maximum parallax factor (or in the case of the WISE data 
points, have large astrometric uncertainties) and do not yet properly constrain the size of the parallactic motion.

Data in Table~\ref{TYparallaxes} are used to plot the absolute magnitude versus spectral type
diagrams shown in Figures~\ref{MH_TYtype} and~\ref{MW2_TYtype}. Figure~\ref{MH_TYtype} shows a sharp drop in the
absolute $H$ magnitude near the T/Y transition. The absolute $H$-band magnitude for our latest object -- the $\ge$Y2
dwarf WISE 1828+2650 -- appears to indicate a stagnation or perhaps even a brightening in the $H$ flux for later
Y dwarfs relative to ones at Y0 and Y1. Figure~\ref{MW2_TYtype} shows an even more surprising result: the absolute
W2 magnitude plummets at Y0 and Y1 only to rebound appreciably for WISE 1828+2650 at $\ge$Y2.

We suggest three possible scenarios to explain this observed behavior. (1) The parallax value for WISE 1828+2650 is the
only one of the Y dwarf parallax measures robust enough to trust. In this case, the plummeting absolute
magnitudes and Y0 and Y1 can be discounted until longer timeline astrometric monitoring of these objects has produced
results with smaller measurement errors.
(2) Despite their large errors, the Y0 and Y1 points indicate a pronounced dimming of the $H$ and, more notably,
W2 fluxes; the discrepant point for WISE 1828+2650 merely demonstrates that it (and not the other Y dwarfs) is the
oddball, its weirdness being attributable to an unknown physical cause or misclassification due to a peculiar
spectral morphology.\footnote{Note that invoking unresolved binarity for this object would account for only 0.75 mag
of brightening, far below what is required to account for the W2 discrepancy.}
(3) All trigonometric parallaxes are credible, and the observed behavior represents a real brightening of
the fluxes for later Y dwarfs relative to ones at Y0-Y1. 

It is clear that continued astrometric monitoring programs targeting Y dwarfs at improved precisions are 
critical in determining which of these scenarios is the 
correct one. For now, we perform weighted least-squares fits to the absolute magnitude versus spectral
type diagrams with the caveat that they represent our best determinations at this time and will certainly
have to be revised as future data become available. 
Third-order polynomials were fit to data in these diagrams both with and without WISE 1828+2650; for fits that
included WISE 1828+2650, a spectral type of exactly Y2 was assumed for the object.
For Figure~\ref{MH_TYtype}, fits included the two upper 
limit M$_H$ values by assuming that the 
limits measured in $H$-band were actual detections with errors equal to the error in the $H$-band measurement of 
WISE 1828 (0.24 mag). Although this is
an {\it ad hoc} assumption, we feel that it is better to fit (conservatively) to available data in that region 
rather than to drop the information entirely.

The resulting least-square fits to the data are given by the following equations:
$$M_H = 20.272231 + 1.9695993(type) + 0.23810003(type)^2 + 0.015161356(type)^3,$$ which includes 
WISE 1828+2650 (red dashed curve in Figure~\ref{MH_TYtype}), or
$$M_H = 21.179419 + 2.9205146(type) + 0.53564210(type)^2 + 0.043616246(type)^3,$$ which excludes it 
(blue solid curve in Figure~\ref{MH_TYtype}). Also,
$$M_{W2} = 14.249389 + 0.60675032(type) + 0.12183371(type)^2 + 0.014785180(type)^3,$$ which includes 
WISE 1828+2650 (red dashed curve in Figure~\ref{MW2_TYtype}), or
$$M_{W2} = 15.213242 + 1.5891486(type) + 0.42455282(type)^2 + 0.043434368(type)^3,$$ which excludes it 
(blue solid curve in Figure~\ref{MW2_TYtype}). In each equation, the spectral type 
is given by $type$ = 0 for Y0, 1 for Y1, $-$1 for T9, $-$2 for T8, etc. These relations are considered valid 
only over the range from T7 to Y1. 

For subsequent analyses we have used only the fits that exclude WISE 1828+2650 because these relations 
are needed solely to estimate distances to objects of earlier type than WISE 1828+2650 itself. In this
case, we are explicitly assuming that scenario (1) above is incorrect and that either (2) or (3) is a
more credible hypothesis. If we instead discover later that this assumption is incorrect, it will have
the effect of increasing the distances to the Y0 and Y1 dwarfs, which will result in a reduction of
the computed space density for objects at those types.

Using these fits, we have estimated distances for each of the objects identified in our all-sky census
of brown dwarfs with types $\ge$T6 (Table~\ref{TY_census_photometry}) using their
measured (or assumed) spectral types and measured $H$ and/or W2 magnitudes. These distance estimates
are shown in Table~\ref{TY_census_distances}. 

We have checked the distance distribution of objects in each spectral type bin by performing the
$V/V_{max}$ test (\citealt{schmidt1968}), which checks the uniformity of a distribution of objects
in space. 
The quantity $V$ is the volume of space interior to object $i$ at distance
$d_i$, and $V_{max}$ is the full volume of space contained within the distance limit, $d_{max}$, of 
the sample. For a uniform sample, the average value, $<V/V_{max}>$, should be 0.5 because
half of the sample should lie in the nearer half of the volume and the rest should lie in the farther half.
If this number is not near 0.5, then the sample is either non-homogeneous or incomplete. 
$V/V_{max}$ values are given in column 8 of Table~\ref{TY_census_distances} for each object falling 
within $d_{max}$. The average value of $V/V_{max}$ in each integral spectral type bin is listed in 
Table~\ref{space_density_numbers}.

For our sample
we find that the T dwarf bins have $<V/V_{max}>$ values less than 0.5. This indicates that objects near 
the $d_{max}$ limit have yet to be identified, meaning that
the sample is still slightly incomplete at larger distances.  
For Y dwarfs (except for the Y2 bin with only a single object), the $<V/V_{max}>$ values are even
further below 0.5. This indicates larger incompleteness at the furthest reaches of the volume, which
is not surprising given that follow-up has so far concentrated on only the closest, brightest examples.
Concentrated work on the discovery of additional Y dwarfs is still badly
needed.

Table~\ref{space_density_numbers} lists the final number of objects in each bin along with the
measured space density for each. 
For the mid- to late-T dwarfs, whose mapping from spectral type to absolute magnitude
is reasonably well established, our two biggest sources of error are sample incompleteness and 
the effects of binarity (\citealt{kirkpatrick2011}). For the former, we can use the $<V/V_{max}>$
values above to estimate the extent of the incompleteness per bin. If we assume that the incompleteness
lies solely in the third of the volume furthest from the Sun, then we can estimate the number of
objects needed to give a value of $<V/V_{max}> = 0.5$. If we arbitrarily place all such missing 
objects at the midpoint of that incomplete shell, each object would have $<V/V_{max}> = 0.84$, which 
corresponds to a distance of 18.9, 14.2, and 9.4 pc, respectively, for $d_{max}$ values of 20, 15, and
10 pc. Under these assumptions, we find that a total of 10, 12, 20, and 4 additional objects are needed to
fill out the T6-6.5, T7-7.5, T8-8.5, and T9-9.5 bins, respectively.

This complete sample, however, will still suffer from the effects of unresolved binarity, the 
brighter magnitude of the composite system leading to a nearer distance estimate than is actually the case.
This will cause a few of the nearest objects to be pushed to larger distances that still fall within
$d_{max}$, but the bigger effect will be in eliminating objects entirely from the sample that were just
within the $d_{max}$ limit before correction. \cite{burgasser2007} summarize the results of high-resolution
imaging and radial velocity surveys for very low mass objects in the field along with high-resolution
imaging surveys of very low mass objects in young clusters and conclude that as many as 30\% (or more)
of these objects could be binary. As a worst-case scenario, we 
assume that 30\% of our systems are eliminated as having distances
beyong $d_{max}$ once binarity is taken into account. We then find that the sample numbers previously corrected
for incompleteness should now be reduced by 14, 13, 17, and 8 objects in the T6-6.5, T7-7.5, T8-8.5, and 
T9-9.5 bins, respectively.

For the Y dwarf bins, the uncertain mapping from spectral type to absolute magnitude along
with known incompleteness of the sample are likely the biggest contributors to the uncertainty in the
space density estimates. We consider the sample incompleteness to be the dominant effect, and hence we
believe that our measured space density estimates for the Y dwarfs should be considered lower limits.

Results are plotted in Figure~\ref{space_density1}. Also
plotted are the results of 
luminosity function simulations by \cite{burgasser2007-newMonteCarlo} 
(based on earlier work by \citealt{burgasser2004}) that show the expected distribution
of objects for power-law mass functions with various slopes of $\alpha$, the functional form of which
is given by $dN/dM \propto M^{-\alpha}$, where $N$ is number of objects and $M$ is the mass.
A nominal low-mass cutoff of 1 $M_{Jupiter}$, a constant birthrate over 0.1-10 Gyr, and 
evolutionary models of \cite{baraffe2003} are
assumed for the simulations, as well as an overall normalization of 0.0037 objects pc$^{-3}$ for stars
with 0.09-0.1 M$_\odot$, based on \cite{reid1999}. We show results for $\alpha = -1, 0, +1$ as well
as the effect of increasing the low-mass cutoff to 5 $M_{Jupiter}$ or
to 10 $M_{Jupiter}$ for the $\alpha = 0$ model. Our measurements, shown by the purple line, are
overplotted on the simulations for comparison.

It is obvious from this diagram that the density of late-T dwarfs falls most closely along
the $\alpha = -1.0$ model. For these results to match more closely to the $\alpha = 0.0$ model
would require that our current census of these objects be deficient by a factor of 2 to 3,
which does not seem plausible given that the follow-up presented in \cite{kirkpatrick2011} and
Mace et al.\ (in prep.) has already completed a large portion of the follow-up needed for the 
brightest WISE T dwarf candidates over the entire sky. Although densities as large as those
predicted by an intermediate $\alpha = -0.5$ model are still possible, these brown dwarfs appear to be
rarer than the $\alpha = 0.0$ simulation predicts.

Turning to the Y dwarfs, we find that the number density climbs relative to the T dwarfs.
This would ordinarily suggest a steeper slope ($\alpha > 0.0$), if these fits were
done independently of results at higher masses. The simulations of \cite{burgasser2004,burgasser2007} are normalized
so that the space density for low-mass stars matches observational measures, so the space densities in Figure~\ref{space_density1} 
are what would be expected if the same power law applied from low-mass stars to low-mass
brown dwarfs. It may simply be that the power-law approximation is not ideal at lower masses, or 
it may be that our space densities for Y dwarfs are overestimates if our trigonomeric parallaxes
for the Y0 and Y1 dwarfs are systematically too large. If we use the red curves from Figures~\ref{MH_TYtype}
and~\ref{MW2_TYtype} (i.e., the relations including WISE 1828+2650 in the fits), we find that the number
densities in the Y dwarf bins drop by a factor of three. At face value this would result in
space densities that more closely favor a slope of $\alpha = 0.0$ (but not its normalization).
However, it must be kept in mind that the coldest bins in Figure~\ref{space_density1} are based on 
only a handful of discoveries identified so far. Other Y dwarfs are certain to be added as 
other candidates objects from WISE are more fully characterized and as Y dwarf companions to
higher mass objects are uncovered. Thus, we expect that these pessimistic values of the Y dwarf
space density themselves represent only lower limits. 

Models by \cite{burrows2003} show that a 
brown dwarf of mass 5 $M_{Jupiter}$ takes $\sim$5 Gyr to cool to 200K; a brown dwarf of
mass 10 $M_{Jupiter}$, even if formed at the same time as the Milky Way itself, has not yet had enough time
to cool to 200K. Our preliminary results
suggest that the low-mass cutoff for star formation is below 10 and possibly even below 5
$M_{Jupiter}$ (see Figure~\ref{space_density1}).
Old brown dwarfs with masses below 5 $M_{Jupiter}$ fall
into a temperature regime ($T_{eff} <$ 200K) that is too faint for WISE to sample a sufficient
volume. As such, we may not be able to place stronger constraints
on the low-mass cutoff using WISE discoveries alone.

It is possible to produce a larger number of cold Y dwarfs relative to L and T dwarfs -- and
therefore provide a better match to our results -- if we drop the assumption that the star
formation rate has been constant over the last 10 Gyr and instead use an exponentially declining 
star formation rate over time or a scenario whereby all star formation took place in the first 
1 Gyr of the Milky Way's lifetime. \cite{burgasser2004} confirm that the numbers of cold brown
dwarfs is markedly higher than that of warmer brown dwarfs under both of these assumptions, but he
cautions that there is little other physical evidence supporting star formation rates like these.
There are other possibilities, too. The \cite{burgasser2004} simulations do not account for 
relative differences in the
dynamical ``heating'' of low-mass versus high-mass brown dwarfs over time. That is, the scale height of the
lower mass objects is expected to be larger than that of the higher mass group, due to the
greater influence on boosting the velocities of the lower mass brown dwarfs during encounters with other objects as they
orbit the Galaxy. However, this would tend to make the low-mass brown dwarfs somewhat rarer in the Solar Neighborhood,
making the mismatch between our results and the simulations even more striking.
 
With these caveats in mind, we compare these preliminary results to other work in the literature
that has attempted to measure the substellar mass function. We select four such avenues of
exploration to use as comparison: earlier wide-field surveys for field brown dwarfs, 
a deep search using the {\it Spitzer Space Telescope}, studies
in young clusters and star formation regions, and the results of field objects indirectly
detected by microlensing surveys.

\subsubsection{Comparison to Results from other Field Brown Dwarf Studies}
A few other wide-field searches have earlier attempted to determine the slope of the field mass function for objects later than
spectral type T6. \cite{metchev2008} used a cross-correlation of 2MASS with the SDSS Data Release 1
to determine a space density of $4.7^{+3.1}_{-2.8}{\times}10^{-3}$ objects pc$^{-3}$ over the range T6-T8.
Their results favor $\alpha \approx 0$. \cite{reyle2010} used data from the Canada-France Brown Dwarf
Survey to determine space densities of $5.3^{+3.1}_{-2.2}{\times}10^{-3}$ objects pc$^{-3}$ for T6-T8
and $8.3^{+9.0}_{-5.1}{\times}10^{-3}$ objects pc$^{-3}$ for T8.5-T9. Their results favor $\alpha \la 0$.
\cite{burningham2010} used data from UKIDSS to determine space densities in the range 
$0.30{\pm}0.20 {\times}10^{-3}$ to $0.59{\pm}0.39 {\times}10^{-3}$ objects pc$^{-3}$ for T6-T6.5,
$0.40{\pm}0.28 {\times}10^{-3}$ to $0.79{\pm}0.55 {\times}10^{-3}$ objects pc$^{-3}$ for T7-T7.5,
$0.58{\pm}0.51 {\times}10^{-3}$ to $1.1{\pm}1.0 {\times}10^{-3}$ objects pc$^{-3}$ for T8-T8.5, and
$3.1{\pm}2.9 {\times}10^{-3}$ to $7.6{\pm}6.9 {\times}10^{-3}$ objects pc$^{-3}$ for T9. These results favor $\alpha < 0$,
confirming the earlier UKIDSS analysis by \cite{pinfield2008}. Figure~\ref{space_density2} summarizes these results
for direct comparison to our results in Figure~\ref{space_density1}.

It should be noted that results from each of these studies were derived using a relatively small number of sources --
4 objects with types of T6 or later for \cite{metchev2008}, 13 objects with types $\ge$T6 for
\cite{reyle2010}, and 25 objects with types $\ge$T6 for \cite{burningham2010}. Our new WISE study
uses a sample of 148 objects (Table~\ref{space_density_numbers}) with types $\ge$T6, almost six times 
the number in the most data-rich of
these previous surveys. It should be noted that the average distance to the \cite{reyle2010} and \cite{burningham2010} samples
are much larger than for the sample we present in Table~\ref{TY_census_photometry} and 
Table~\ref{TY_census_distances}. Our WISE-enabled sample is
the nearest sample possible, and thus the one for which parallaxes and info on unresolved
binarity will be most readily obtained, allowing for refinements of our space density 
numbers to unequalled precisions in the future. Nevertheless, despite the sparseness of the earlier data,
those results are in general agreement with our space density results for objects earlier than T9.

\subsubsection{Comparison to Results from the {\it Spitzer} Deep, Wide-field Survey}

\cite{eisenhardt2010} identified a sample of 14 cool brown dwarf candidates (spectral types $>$T7) 
using data from the 10-deg-square {\it Spitzer} Deep, Wide-field Survey (SDWFS).
The coldest of these candidates, SDWFS J143356.62+351949.2, has colors of ch1$-$ch2=2.24$\pm$0.46 mag 
and $H-$ch2$>$5.73 mag. Using figures 11 and 14 of \cite{kirkpatrick2011} we find that the ch1$-$ch2 
color is typical of a T9 dwarf and the $H-$ch2 color suggests that its type is $\ge$T8. This survey,
therefore, was probing down as far as late-T but probably not to early-Y.

\cite{eisenhardt2010} found that the assumed space density and measured color distribution of the 
14 brown dwarf candidates most closely matched a \cite{chabrier2003} log-normal mass function, or 
a power-law mass function with $\alpha=1.3$, as long as the 4.5-$\mu$m flux of the \cite{burrows2003} 
models was adjusted downward\footnote{Such flux suppression is expected if non-equilibrium chemistry 
is altering the predicted depths of the CO fundamental band at 
4.7 $\mu$m (\citealt{golimowski2004}, \citealt{saumon2007}).}. This value of $\alpha$, however, is very high compared to the one we deduce from 
our measurements of field, late-T dwarfs in the all-sky sample ($-0.5 < \alpha < 0.0$).

New data has become available that sheds additional light on these results. 
SDWFS J143356.62+351949.2 was imaged with {\it HST}/WFC3 in eight exposures of 1400s duration on 
2009 Dec 03, and in twelve exposures of 700s duration on 2009 Dec 16, as part of program 12044.  
Assuming a magnitude of W2 $\approx$ ch2 = 18.47 mag, the marginal F140W detection at 26.2 mag 
is fainter than the $\sim$25 mag expected for a T9 dwarf (see Figure~\ref{YDwarfs_Colors}).
The F127M filter  covers 1.239 to 1.308 $\mu$m, and from Figure~\ref{YDwarfs2} and Figure~\ref{YDwarfs_Jcomparison} should 
include at least half the light seen in the 1.193 to 1.592 $\mu$m  F140W filter.  With $<$20\% of 
the F140W bandpass, the F127M filter should thus produce a noticeably brighter magnitude if the
object were a cold brown dwarf.  However, the source is completely undetected in F127M, with 2$\sigma$ limit 
of 26.2 mag. We consider it more likely that this object is an extremely red extragalactic source.
The SDWFS results, because of their depth, may have been more heavily contaminated by
extragalactic objects than previously believed, leading to the higher value of $\alpha$.

\subsubsection{Comparison to the Initial Mass Function in Open Clusters and Star-Forming Regions}

Because of it proximity, relatively young age, and compactness on the sky, the Pleiades has been 
one of the most popular open clusters in which to hunt for brown dwarfs. Many studies have attempted 
to measure the slope of the substellar initial mass function in this cluster, with a sampling of 
results suggesting $\alpha = 0 - 1$ (\citealt{festin1998}), $\alpha = 0.6$ (\citealt{bouvier1998}), 
and $\alpha = 0.62{\pm}0.14$ (\citealt{casewell2007}). Other open clusters are found to have initial 
mass functions that are similar to that of the Pleiades (\citealt{bastian2010}). These values of
$\alpha$ are slightly larger than the one we derive.

Studies of the brown dwarf population in star-formation regions is hampered by the larger distances 
to these areas and to complications with survey completeness, which is impacted by variable extinction 
across the field. However, low-mass brown dwarfs are still in a young, warm state and are easier 
to detect than in older clusters. \cite{luhman2007-cluster} compare results in Taurus, Chamaeleon I, IC 348, 
and the Trapezium Cluster to conclude that an initial mass function with slope $\alpha \lesssim 0$ is 
consistent with the ratio of the number of stars to brown dwarfs seen in those regions. Although 
doubts about the universality of the initial mass function persist, \cite{andersen2008} found that 
this assumption is credible and, having combined observational results in seven star formation 
regions and open clusters in the mass range 0.03-1.0 M$_\odot$ -- including the Pleiades, the Orion Nebula Cluster, and Taurus -- 
conclude that $\alpha < 0$. These studies are in good agreement with our results using field
brown dwarfs.

\subsubsection{Comparison to the Number Density of Solivagant Objects found by Microlensing Surveys}

\cite{sumi2011} have used the collection of observed microlensing events detected in the 
Microlensing Observations in Astrophysics (MOA) and Optical Gravitational Lensing Experiment 
(OGLE) surveys to study the random stellar and substellar populations of the Milky Way. These 
populations are indirectly observed as lensing sources that temporarily brighten background 
sources in the Galactic Bulge as they pass through the line of sight between the earth and the 
Bulge source. The duration of the magnification event depends upon the mass, distance, and transverse 
speed of the lens, and can be related back to mass using assumptions about the kinematic and 
space distribution of the lenses. \cite{sumi2011} identify 474 well characterized events with 
which they study the field mass function of the lensing sources. They fit these results with a 
tripartite power-law function that, in the brown dwarf regime ($0.01 \le$ $M/M_\sun \le 0.08$), 
has an exponent of $\alpha = 0.49^{+0.24}_{-0.27}$. This value of $\alpha$ is similar to that
found for brown dwarfs in the Pleiades but is somewhat larger than what we find in the field.

\cite{sumi2011} also find that another population -- rogue planets -- may be needed to explain 
results at very low masses. However, this population is detectable over the low-mass tail of the 
brown dwarf population only at masses below $\sim$4 $M_{Jupiter}$ (Einstein radius crossing times 
of less than 2 days). Models from \cite{burrows2003} predict that an old (5 Gyr) solar metallicity 
brown dwarf with mass of  5 $M_{Jupiter}$ would have an effective temperature of only 200K and 
would be too dim to be imaged by WISE beyond $\sim$1 pc. Given that these objects would be much 
rarer than low-mass brown dwarfs (see figure 2 of \citealt{sumi2011}), they will not have any 
impact on the WISE results presented here.

\section{Conclusions}

We have presented seven new Y dwarf discoveries from WISE, bringing the total number of Y dwarf
discovered by WISE to thirteen. Using these Y dwarf discoveries along with WISE T dwarf discoveries
from \cite{kirkpatrick2011} and Mace et al.\ (in prep.) and discoveries from previous searches for
field brown dwarfs, we compute space densities for late-T and early-Y dwarfs and find that
stars outnumber brown dwarfs in the Solar Neighborhood by a factor of roughly six. However, this
factor is certain to shrink in the future for two reasons: (1) the census of stars within a few
parsecs of the Sun is well known, whereas that for late-T and Y dwarfs is still incomplete, so the
ratio of stars to brown dwarfs is expected to decrease in the future; (2) the field sample does
not yet show any clear signature that we have probed beyond star formation's low-mass cutoff, meaning 
that colder, very low-mass objects likely exist and are too faint for even WISE to detect. This
having been said, however, we expect the total number of Y dwarfs identified by WISE to be only a few
dozen when follow-up is largely complete. Unless there is a {\it vast} reservoir of cold brown dwarfs
invisible to WISE, the space density of stars is still expected to greatly outnumber that of brown dwarfs.

Because of the extreme faintness of these Y dwarfs, we have so far only been able to tease out a few
hints regarding their their distances, their atmospheric details, and their
variety. Future work is needed to identify more examples and to characterize more fully the physical 
properties of ones already recognized. The study of Y-type brown dwarfs is still in its infancy
but the glimpses caught so far promise tantalizing results ahead.

\section{Acknowledgments}

We thank the referee, Sandy Leggett, for constructive comments that helped to improve the paper.
This publication makes use of data products from the Wide-field Infrared Survey Explorer, which is a 
joint project of the University of California, Los Angeles, and the Jet Propulsion Laboratory/California 
Institute of Technology, funded by the National Aeronautics and Space Administration. 
This publication also makes use of data products from 2MASS, SDSS, and DSS. 
2MASS is a joint project of the University of Massachusetts and the Infrared Processing and Analysis 
Center/California Institute of Technology, funded by the National Aeronautics and Space Administration 
and the National Science Foundation. 
SDSS is funded by the Alfred P. Sloan Foundation, the Participating 
Institutions, the National Science Foundation, the U.S. Department of Energy, the National Aeronautics 
and Space Administration, the Japanese Monbukagakusho, the Max Planck Society, and the Higher Education 
Funding Council for England.
The DSS were produced at the Space Telescope Science Institute under
U.S.\ Government grant NAG W-2166. The images of these surveys are based on photographic data
obtained using the Oschin Schmidt Telescope on Palomar Mountain and the UK Schmidt Telescope.

This work is based in part on observations made with the {\it Spitzer} Space Telescope, which is
operated by the Jet Propulsion Laboratory, California Institute of Technology, under a contract with
NASA. Support for this work was provided by NASA through an award issued to program 70062 and 80109 by JPL/Caltech. 
This work is also based in part on observations made with the NASA/ESA {\it Hubble} Space Telescope, obtained
at the Space Telescope Science Institute, which is operated by the Association of Universities for
Research in Astronomy, Inc., under NASA contract NAS 5-26555. These observations are associated with 
programs \#12044 and \#12330. Support for these programs was provided by NASA through a grant from the Space
Telescope Science Institute.
Some of the spectroscopic data presented herein were obtained at 
the W.M. Keck Observatory, which is operated as a scientific partnership among 
the California Institute of Technology, the University of California and the 
National Aeronautics and Space Administration. The Observatory was made 
possible by the generous financial support of the W.M. Keck Foundation.
In acknowledgement of our observing time at Keck and the IRTF, 
we further wish to recognize the very significant 
cultural role and reverence that the summit of Mauna Kea has always had within the indigenous Hawai'ian 
community. We are most fortunate to have the opportunity to conduct observations from this mountain.  

Our research has benefitted from the M, L, and
T dwarf compendium housed at DwarfArchives.org, whose server was funded by a NASA Small Research Grant, 
administered by the American Astronomical Society. 
This research has made use of the NASA/IPAC Infrared Science Archive (IRSA),
which is operated by the Jet Propulsion Laboratory, California Institute of Technology, under contract
with the National Aeronautics and Space Administration. 
We are also indebted to the SIMBAD database,
operated at CDS, Strasbourg, France.

\clearpage

\begin{deluxetable}{cccccccc}
\tabletypesize{\tiny}
\tablewidth{6in}
\tablecaption{WISE All-Sky Release Photometry of WISE Y Dwarfs\label{y_dwarf_discoveries_wise}}
\tablehead{
\colhead{WISE Designation} &                          
\colhead{W1} &        
\colhead{W2} &              
\colhead{W3} &              
\colhead{W4} &       
\colhead{W1-W2} &    
\colhead{W2-W3} &
\colhead{Spec.} \\         
\colhead{} &                          
\colhead{(mag)} &        
\colhead{(mag)} &              
\colhead{(mag)} &              
\colhead{(mag)} &       
\colhead{(mag)} &    
\colhead{(mag)} &
\colhead{Type}\\        
\colhead{(1)} &                          
\colhead{(2)} &   
\colhead{(3)} &       
\colhead{(4)} &         
\colhead{(5)} &        
\colhead{(6)} &              
\colhead{(7)} &              
\colhead{(8)} \\
}
\startdata
WISE J014656.66+423410.0                     & $>$18.99      & 15.08$\pm$0.07& 13.20$\pm$0.51& $>$9.41      & $>$3.91      & 1.88$\pm$0.51 & Y0      \\
WISE J035000.32$-$565830.2                   & $>$18.90      & 14.73$\pm$0.06& 12.53$\pm$0.29& $>$9.00      & $>$4.17      & 2.20$\pm$0.30 & Y1      \\      
WISE J035934.06$-$540154.6                   & $>$19.29      & 15.42$\pm$0.07& $>$13.02      & $>$9.76      & $>$3.87      & $<$2.40       & Y0      \\     
WISE J041022.71+150248.4\tablenotemark{a}    & $>$18.33      & 14.18$\pm$0.06& $>$11.86      & $>$8.90      & $>$4.15      & $<$2.32       & Y0      \\
WISE J053516.80$-$750024.9                   & $>$18.92      & 15.06$\pm$0.07& 13.12$\pm$0.45& $>$9.00      & $>$3.86      & 1.94$\pm$0.46 & $\ge$Y1:\\
WISE J071322.55$-$291751.9                   & $>$18.35      & 14.48$\pm$0.06& 12.67$\pm$0.43& $>$9.22      & $>$3.87      & 1.81$\pm$0.43 & Y0      \\
WISE J073444.02$-$715744.0                   & $>$19.42      & 15.36$\pm$0.06& 13.65$\pm$0.42& 9.65$\pm$0.33& $>$4.06      & 1.71$\pm$0.42 & Y0      \\
WISE J140518.39+553421.3\tablenotemark{a}    & $>$18.82      & 14.10$\pm$0.04& 12.43$\pm$0.27& $>$9.40      & $>$4.72      & 1.67$\pm$0.27 & Y0 pec? \\
\enddata
\tablenotetext{a}{For details about these sources, please see \cite{cushing2011}.}
\tablenotetext{b}{This source has a bluer W1-W2 color than it did in earlier WISE processing -- W1-W2=2.49$\pm$0.18 now versus W1-W2$>$3.04 before 
(\citealt{kirkpatrick2011}). In both cases, the strong W2 detection would have driven the photometry, but in the newer processing, the code appears 
to have erroneously assigned to this object some of the W1 flux from the bright
star to the north, therefore producing a brighter W1 detection for WISE 1541$-$2250 than actually exists.}
\end{deluxetable}

\clearpage

\begin{deluxetable}{lccccccccccc}
\tabletypesize{\tiny}
\rotate
\tablewidth{8.5in}
\tablenum{2}
\tablecaption{Photometric Follow-up of Y Dwarfs\label{y_dwarf_discoveries_phot}}
\tablehead{
\colhead{WISE Designation} &
\colhead{F140W} &                          
\colhead{$J$} &              
\colhead{$H$} & 
\colhead{Ref} &            
\colhead{ch1} &              
\colhead{ch2} &             
\colhead{ch1-ch2} &
\colhead{F140W-W2} &
\colhead{$J$-W2} &
\colhead{$H$-W2} &
\colhead{Spec.} \\
\colhead{} &                          
\colhead{(mag)} &              
\colhead{(mag)} &             
\colhead{(mag)} & 
\colhead{} &             
\colhead{(mag)} &             
\colhead{(mag)} &             
\colhead{(mag)} &              
\colhead{(mag)} &             
\colhead{(mag)} &              
\colhead{(mag)} &
\colhead{Type} \\
\colhead{(1)} &                          
\colhead{(2)} &   
\colhead{(3)} &       
\colhead{(4)} &         
\colhead{(5)} &        
\colhead{(6)} &              
\colhead{(7)} &              
\colhead{(8)} &       
\colhead{(9)} &    
\colhead{(10)} &
\colhead{(11)} &
\colhead{(12)} \\
}
\startdata
WISE J014656.66+423410.0   & ---           & {\it 19.40$\pm$0.25}           & {\it 18.71$\pm$0.24}           &  1&  17.50$\pm$0.07& 15.08$\pm$0.02& 2.42$\pm$0.07& ---          & {\it 4.32$\pm$0.26}& {\it 3.63$\pm$0.25}& Y0      \\ 
WISE J035000.32$-$565830.2 & 22.30$\pm$0.20& $>$22.8                        & $>$21.5                        &  2&  17.94$\pm$0.10& 14.69$\pm$0.02& 3.25$\pm$0.10& 7.57$\pm$0.21& $>$8.1             & $>$6.8             & Y1      \\      
WISE J035934.06$-$540154.6 & 21.84$\pm$0.20& 21.56$\pm$0.24                 & 22.20$\pm$0.43                 &  2&  17.55$\pm$0.07& 15.33$\pm$0.02& 2.22$\pm$0.08& 6.42$\pm$0.21& 6.14$\pm$0.25      & 6.78$\pm$0.44      & Y0      \\     
WISE J041022.71+150248.4   & ---           & 19.24$\pm$0.05\tablenotemark{a}& 19.05$\pm$0.09                 &  3&  16.64$\pm$0.04& 14.18$\pm$0.02& 2.46$\pm$0.05& ---          & 5.06$\pm$0.08      & 4.87$\pm$0.11      & Y0      \\
WISE J053516.80$-$750024.9 & 22.42$\pm$0.20& $>$21.1                        & $>$21.6                        &  2&  17.77$\pm$0.09& 15.01$\pm$0.02& 2.76$\pm$0.09& 7.36$\pm$0.21& $>$6.0             & $>$6.5             & $\ge$Y1:\\
WISE J071322.55$-$291751.9 & ---           & 19.64$\pm$0.15                 & $>$19.3                        &  4&  16.67$\pm$0.04& 14.22$\pm$0.02& 2.45$\pm$0.05& ---          & 5.16$\pm$0.16      & $>$4.8             & Y0      \\ 
                           & ---           & {\it 20.06$\pm$0.21}           & {\it 21.16$\pm$0.66}           & 10&  ---           & ---           & ---          & ---          & {\it 5.58$\pm$0.22}& {\it 6.68$\pm$0.66}& ---     \\
WISE J073444.02$-$715744.0 & ---           & 20.41$\pm$0.27                 & ---                            &  5&  17.69$\pm$0.08& 15.21$\pm$0.02& 2.47$\pm$0.08& ---          & 5.05$\pm$0.28      & ---                & Y0      \\
\enddata   
\tablecomments{All $J$ and $H$ data are on the MKO-NIR filter system except for those in italics. See text for details.}
\tablecomments{References for photometry: 
(1) Bigelow/2MASS on 2010 Nov 20 UT; 
(2) Magellan/PANIC on 2010 Nov 25 UT except for the $J$-band measurement of WISE 0359$-$5401, which was taken on 2010 Aug 01 UT; 
(3) Palomar/WIRC on 2010 Jul 26 except for the $J$ measurement of WISE 0410+1502, which was taken on 2010 Aug 29 UT; 
(4) SOAR/Spartan-IRC on 2012 Jan 13 UT; 
(5) CTIO4m/NEWFIRM on 2011 Apr 17 UT; 
(6) Palomar/WIRC combined -- $J$ is from 2010 Jul 26 UT and $H$ is from a mosaic made from data taken on 2010 Jul 26, Aug 28, Aug 30, and Aug 31 UT; 
(7) Keck/NIRC2+LGSAO on 2010 Jul 01 UT; 
(8) Palomar/WIRC 2010 Aug 29 UT for $J$ and 2011 May 16 UT for $H$; 
(9) AAT/IRIS2 on 2011 Sep 12 UT for $J$ and SOAR/OSIRIS on 2011 Sep 18 UT for $H$; 
(10) SOAR/OSIRIS on 2012 Mar 13 UT for $J$ and 2012 Mar 12 UT for $H$.}
\tablenotetext{a}{These values have been updated using of an aperture with radius set to 1.5 times 
the full width at half maximum of the observed point spread function, as opposed to the
use of a standard, fixed aperture size for all observations, as had been done in \cite{cushing2011}.}
\end{deluxetable}

\clearpage

\begin{deluxetable}{ccccccccccc}
\tabletypesize{\tiny}
\tablewidth{4.5in}
\tablenum{3}
\tablecaption{New Follow-up Spectroscopy of Y Dwarfs\label{y_dwarf_discoveries_spec}}
\tablehead{
\colhead{WISE Designation} &                          
\colhead{Instrument} &       
\colhead{Obs.\ date} &         
\colhead{Int.\ Time} &  
\colhead{Spec.} \\ 
\colhead{} &                          
\colhead{} &        
\colhead{(UT)} &
\colhead{(sec)} &              
\colhead{Type} \\        
\colhead{(1)} &                          
\colhead{(2)} &   
\colhead{(3)} &       
\colhead{(4)} &         
\colhead{(5)} \\
}
\startdata
WISE J014656.66+423410.0    & Keck/NIRSPEC-N3           & 2011 Aug 08       &  6000                  &  Y0      \\
                            & Keck/NIRSPEC-N5           & 2011 Sep 06       &  3600                  &  Y0      \\
WISE J035000.32$-$565830.2  & HST/WFC3                  & 2011 Aug 13       &  2212                  &  Y1      \\      
WISE J035934.06$-$540154.6  & HST/WFC3                  & 2011 Aug 09-10    &  2212                  &  Y0      \\     
WISE J053516.80$-$750024.9  & HST/WFC3\tablenotemark{a} & 2011 Sep 27       &  2212                  &  $\ge$Y1:\\
WISE J071322.55$-$291751.9  & Keck/NIRSPEC-N3           & 2011 Oct 07-09    & 12000                  &  Y0      \\
WISE J073444.02$-$715744.0  & Magellan/FIRE             & 2012 Jan 05       &  1775                  &  Y0      \\
WISE J205628.91+145953.2    & Keck/NIRSPEC-N3           & 2011 Aug 09       &  1200\tablenotemark{b} &  Y0      \\
\enddata
\tablenotetext{a}{This spectrum suffers heavy contamination from a field object.}
\tablenotetext{b}{This spectrum was added to the 2400s NIRSPEC-N3 spectrum previously reported in \cite{kirkpatrick2011}.}
\end{deluxetable}

\clearpage

\begin{deluxetable}{lllllllllll}
\tabletypesize{\tiny}
\rotate
\tablewidth{8.5in}
\tablenum{4}
\tablecaption{The Census of Stars and Brown Dwarfs within 8 Parsecs of the Sun\label{8pc_census}}
\tablehead{
\colhead{Discovery Name} &
\colhead{Other Name} &
\colhead{Gliese Name} &
\colhead{$\pi_{trig}$\tablenotemark{a}} &
\colhead{$\sigma_{\pi}$} &
\colhead{$\pi$ Ref.} &
\colhead{Spec.} &
\colhead{Spec.} &
\colhead{J2000 Coords.} &
\colhead{Sys.} &
\colhead{Indiv.} \\
\colhead{} &
\colhead{} &
\colhead{} &
\colhead{(mas)} &
\colhead{(mas)} &
\colhead{} &
\colhead{Type} &
\colhead{Ref.} &
\colhead{} &
\colhead{Rank} &
\colhead{Rank} \\
\colhead{(1)} &
\colhead{(2)} &
\colhead{(3)} &
\colhead{(4)} &
\colhead{(5)} &
\colhead{(6)} &
\colhead{(7)} &
\colhead{(8)} &
\colhead{(9)} &
\colhead{(10)} &
\colhead{(11)} \\
}
\startdata
Sun                          & Sol                 & ---           &     ---&   ---& --&     G2 V      & 51&      ---      &    0 &    0 \\
Proxima Cen\tablenotemark{b} & $\alpha$ Cen C      & Gl 551        &  769.91&  0.54& 16&     M5.5 V    & 18&      1429$-$6240&    1 &    1 \\  
Rigel Kentaurus A            & $\alpha$ Cen A      & Gl 559 A      &  754.81&  4.11&  5&     G2 V      & 50&      1439$-$6050&    1 &    2 \\
Rigel Kentaurus B            & $\alpha$ Cen B      & Gl 559 B      &     ---&   ---& --&     K2 IV C2+1& 50&      1439$-$6050&    1 &    2 \\
Barnard's Star               & ---                 & Gl 699        &  545.4 &  0.3 & 16&     M4 V      & 38&      1757+0441&    2 &    4 \\ 
Wolf 359                     & ---                 & Gl 406        &  419.1 &  2.1 &  6&     M6 V      & 38&      1056+0700&    3 &    5 \\
Lalande 21185                & ---                 & Gl 411        &  392.64&  0.67&  5&     M2 V      & 38&      1103+3558&    4 &    6 \\
Sirius A                     & $\alpha$ CMa A      & Gl 244 A      &  379.21&  1.58&  5&     A0mA1 Va  & 49&      0645$-$1642&    5 &    7 \\
\enddata
\end{deluxetable}

\clearpage

\begin{deluxetable}{lllllll}
\tabletypesize{\tiny}
\rotate
\tablewidth{8.5in}
\tablenum{5}
\tablecaption{Objects Once Listed as Falling Within 8 pc But Now Believed To Lie More Distant\label{8pc_census_out}}
\tablehead{
\colhead{Name} &
\colhead{$\pi_{trig}$} &
\colhead{$\sigma_{\pi}$} &
\colhead{$\pi$ Ref.} &
\colhead{Spec.} &
\colhead{Spec.} &
\colhead{J2000 Coords.}\\
\colhead{} &
\colhead{(mas)} &
\colhead{(mas)} &
\colhead{} &
\colhead{Type} &
\colhead{Ref.} &
\colhead{} \\
\colhead{(1)} &
\colhead{(2)} &
\colhead{(3)} &
\colhead{(4)} &
\colhead{(5)} &
\colhead{(6)} &
\colhead{(7)} \\
}
\startdata
Gl 915                                  &  122.27&  1.13&  3 &  DAP   & 10    & 0002$-$4309 \\  
2MASS J00155808$-$1636578\tablenotemark{a}&  123   &   ---& 12 &  M4 V  &  9, 11& 0015$-$1636 \\ 
Gl 54 AB                                &  122.04&  2.42&  1 &  M2.5 V& 13    & 0110$-$6726 \\
Gl 102\tablenotemark{b}                 &  102.4 &  3.7 &  6 &  M4 V  & 14    & 0233+2455 \\ 
Gl 143.1\tablenotemark{c}               &   45.28&  1.97&  1 &  K7 V  & 15    & 0329$-$1140 \\ 
Gl 178                                  &  123.94&  0.17&  1 &  F6 IV-V&16    & 0449+0657 \\
Gl 185 AB                               &  116.59&  1.51&  1 &  K7 V  & 17    & 0502$-$2115 \\
Gl 190                                  &  107.85&  2.10&  1 &  M3.5 V& 14    & 0508$-$1810 \\
\enddata
\end{deluxetable}

\clearpage

\begin{deluxetable}{cccccccccc}
\tabletypesize{\tiny}
\rotate
\tablewidth{8in}
\tablenum{6}
\tablecaption{Late-T and Y Dwarfs with Measured Trigonometric Parallaxes\label{TYparallaxes}}
\tablehead{
\colhead{Discovery Name} &                          
\colhead{Disc.\ Ref.} &       
\colhead{Spec.\ Type} &  
\colhead{Type Ref.} &       
\colhead{W1} &  
\colhead{W2} &
\colhead{H} &
\colhead{H Ref.} &
\colhead{$\pi_{trig}$} &
\colhead{$\pi_{trig}$ Ref.} \\
\colhead{} &                          
\colhead{} &       
\colhead{} &         
\colhead{} &         
\colhead{(mag)} &  
\colhead{(mag)} &
\colhead{(mag)} &
\colhead{} &
\colhead{(arcsec)} &
\colhead{} \\
\colhead{(1)} &                          
\colhead{(2)} &   
\colhead{(3)} &       
\colhead{(4)} &         
\colhead{(5)} &
\colhead{(6)} &   
\colhead{(7)} &       
\colhead{(8)} &         
\colhead{(9)} &
\colhead{(10)} \\
}
\startdata
SDSS J000013.54+255418.6   & 1        & T4.5& 52& ---           &  ---           &  14.73$\pm$0.07&  38    &  0.0708$\pm$0.0019& 39 \\
ULAS J003402.77$-$005206.7 & 2        & T8.5& 32& 17.47$\pm$0.29&  14.50$\pm$0.08&  18.49$\pm$0.04&  2     &  0.0780$\pm$0.0036& 40 \\
HD 3651B (J0039+2115)      & 3        & T7.5&  3& ---           &  ---           &  16.68$\pm$0.04&  3     &  0.0904$\pm$0.0003& 41 \\
CFBDS J005910.90$-$011401.3& 4        & T8.5& 32& 17.07$\pm$0.15&  13.68$\pm$0.04&  18.27$\pm$0.05&  4     &  0.1011$\pm$0.0025& 39 \\
SDSS J020742.48+000056.2   & 5        & T4.5& 52& 16.39$\pm$0.08&  15.07$\pm$0.08&  ---           &  ---   &  0.0293$\pm$0.0040& 40  \\           
2MASSI J0243137$-$245329   & 6        & T6  & 52& 14.67$\pm$0.03&  12.92$\pm$0.03&  15.14$\pm$0.11&  38    &  0.0936$\pm$0.0036& 42 \\
WISEPA J025409.45+022359.1 & 7, 8     & T8  &  7& 15.76$\pm$0.05&  12.74$\pm$0.03&  16.30$\pm$0.01&  51    &  0.166$\pm$0.026  & 43 \\
WISE J035000.32$-$565830.2 & 51       & Y1  & 51& $>$18.90      &  14.73$\pm$0.06&  $>$21.5       &  51    &  0.238$\pm$0.038  & 43 \\
\enddata
\end{deluxetable}

\clearpage

\begin{deluxetable}{cccccccccc}
\tabletypesize{\tiny}
\rotate
\tablewidth{8in}
\tablenum{7}
\tablecaption{Photometry for Objects in the All-sky Late-T and Y Dwarf Census\label{TY_census_photometry}}
\tablehead{
\colhead{Discovery name} &
\colhead{Disc.} &
\colhead{WISE designation\tablenotemark{a}} &
\colhead{Spec.} &
\colhead{Type} &
\colhead{H} &
\colhead{H} &
\colhead{W2} &
\colhead{W1-W2} &
\colhead{H-W2} \\
\colhead{} &
\colhead{ref.} &
\colhead{} &
\colhead{type} &
\colhead{ref.} &
\colhead{(mag)} &
\colhead{ref.} &
\colhead{(mag)} &
\colhead{(mag)} &
\colhead{(mag)} \\
\colhead{(1)} &
\colhead{(2)} &
\colhead{(3)} &
\colhead{(4)} &
\colhead{(5)} &
\colhead{(6)} &
\colhead{(7)} &
\colhead{(8)} &
\colhead{(9)} &
\colhead{(10)} \\
}
\startdata
\\
\multicolumn{10}{c}{T6-T6.5 Dwarfs} \\
\\
WISEPC J000849.76$-$173922.6&   1& WISE J000849.75$-$173922.9 & T6   &  1&    17.77$\pm$0.13&  1&  14.56$\pm$0.07& 1.86$\pm$0.12&  3.22$\pm$0.15\\
2MASS J00345157+0523050     &   4& WISE J003452.02+052307.0   & T6.5 &  5&    15.44$\pm$0.08& 57&  12.52$\pm$0.03& 2.54$\pm$0.05&  2.92$\pm$0.09\\    
WISE J003830.54+840517.7    &   2& WISE J003830.54+840517.7   & T6   &  2&    ---           &  -&  15.34$\pm$0.09& 2.19$\pm$0.20&  ---          \\  
WISE J013525.64+171503.4    &   2& WISE J013525.64+171503.4   & T6   &  2&    ---           &  -&  14.74$\pm$0.07& 2.45$\pm$0.17&  ---          \\    
WISEPA J022105.94+384202.9  &   1& WISE J022105.97+384203.2   & T6.5 &  1&    17.45$\pm$0.15&  1&  14.70$\pm$0.08& 2.18$\pm$0.17&  2.76$\pm$0.17\\
\enddata
\end{deluxetable}

\clearpage

\begin{deluxetable}{cccccccc}
\tabletypesize{\tiny}
\tablewidth{6in}
\tablenum{8}
\tablecaption{Distances for Objects in Table~\ref{TY_census_photometry}\label{TY_census_distances}}
\tablehead{
\colhead{Discovery name} &
\colhead{Spec.} &
\colhead{${\pi}_{trig}$} &
\colhead{${\pi}_{trig}$} &
\colhead{D$_{H}$} &
\colhead{D$_{W2}$} &
\colhead{D$_{adopted}$} &
\colhead{$\frac{V}{V_{max}}$} \\
\colhead{} &
\colhead{type} &
\colhead{(pc)} &
\colhead{ref.} &
\colhead{(pc)} &
\colhead{(pc)} &
\colhead{(pc)} &
\colhead{} \\
\colhead{(1)} &
\colhead{(2)} &
\colhead{(3)} &
\colhead{(4)} &
\colhead{(5)} &
\colhead{(6)} &
\colhead{(7)} &
\colhead{(8)} \\
}
\startdata
\\
\multicolumn{7}{c}{T6-T6.5 Dwarfs (distance limit = 20 pc, $\langle \frac{V}{V_{max}} \rangle = 0.41$)} \\
\\                                                                                  
WISEPC J000849.76$-$173922.6& T6   & --               & --& 31.5&  21.8&  26.7&  ---\\
2MASS J00345157+0523050     & T6.5 & --               & --&  9.1&   8.1&   8.6& 0.08\\
WISE J003830.54+840517.7    & T6   & --               & --&  ---&  31.2&  31.2&  ---\\
WISE J013525.64+171503.4    & T6   & --               & --&  ---&  23.7&  23.7&  ---\\
WISEPA J022105.94+384202.9  & T6.5 & --               & --& 22.9&  22.0&  22.4&  ---\\
\enddata
\tablecomments{Parallax references:  
(1) \cite{marocco2010},  
(2) \cite{lucas2010},
(3) \cite{artigau2010},
(4) \cite{perryman1997},
(5) \cite{vrba2004},
(6) \cite{vanleeuwen2007},
(7) \cite{schilbach2009}, 
(8) \cite{dahn2002}, 
(9) \cite{tinney2003},
(10) \cite{henry2006},
(11) \cite{dupuy2012},
(12) \cite{kirkpatrick2011},
(13) Beichman et al.\ (in prep.),
(14) Marsh et al.\ (in prep.),
(15) \cite{burgasser2008},
(16)  \cite{subasavage2009}
(17) \cite{vanaltena1995},
(18) \cite{karatas2004}.
}
\tablenotetext{a}{This object may not exist.}
\tablenotetext{b}{Distance estimate from \cite{cuby1999}.}
\tablenotetext{c}{\cite{geissler2011} estimates that the primary (if this is a true binary) has a type of T2. The $J$-band combined light of the system is 16.05 mag from \cite{chiu2006}. The $J$-band absolute magnitude relation from \cite{looper2008} gives $M_J = 14.45$ mag for a T2 dwarf, meaning that this system is no closer than 20.9 pc.}
\tablenotetext{d}{Distance estimate from \cite{lodieu2009}.}
\tablenotetext{e}{Distance estimate from \cite{stumpf2010}.}
\tablenotetext{f}{Distance estimate from \cite{burgasser2012}.}
\tablenotetext{g}{Assumed distance is that derived from the W2 estimate only because the $H$-band measurement is very uncertain.}
\tablenotetext{h}{Assumed distance is that derived from the $H$ estimate only because the W2 measurement is blended with a nearby source.}
\tablenotetext{i}{Distance estimate from \cite{masters2012}.}
\end{deluxetable}

\clearpage

\begin{deluxetable}{ccccccc}
\tabletypesize{\tiny}
\tablewidth{6.5in}
\tablenum{9}
\tablecaption{Updated Space Densities for the All-sky, Volume-limited Sample of Late-T and Y Dwarfs\label{space_density_numbers}}
\tablehead{
\colhead{Spectral Type} &
\colhead{Approx.\ $T_{eff}$} &
\colhead{$d_{max}$} &
\colhead{Total \#}&
\colhead{Obs.\ Space Density} &
\colhead{$<V/V_{max}>$} &
\colhead{\# per 100K Bin} \\
\colhead{Range} &
\colhead{Range (K)} &
\colhead{(pc)} &
\colhead{objects}&
\colhead{(\#/$pc^3$)} &
\colhead{} &
\colhead{w/in 10 pc} \\
\colhead{(1)} &
\colhead{(2)} &
\colhead{(3)} &
\colhead{(4)} &
\colhead{(5)} &
\colhead{(6)} &
\colhead{(7)} 
}
\startdata
T6-T6.5 &  900-1050 & 20 & 38 & 1.1e-3 & 0.41 & 3.1 \\
T7-T7.5 &   750-900 & 20 & 31 & 9.3e-4 & 0.37 & 2.6 \\
T8-T8.5 &   600-750 & 20 & 46 & 1.4e-3 & 0.35 & 3.9 \\
T9-T9.5 &   450-600 & 15 & 22\tablenotemark{a} & 1.6e-3\tablenotemark{a} & 0.44 & 4.5\tablenotemark{a} \\
Y0-Y0.5 &   300-450 & 10 &  8\tablenotemark{a} & 1.9e-3\tablenotemark{a} & 0.30 & 5.3\tablenotemark{a} \\
$\ge$Y1 &   150-300 & 10 &  3\tablenotemark{a} & 7.2e-4\tablenotemark{a} & 0.28 & 2.0\tablenotemark{a} \\
\enddata
\tablenotetext{a}{These values should be taken as lower limits only, given that follow-up of $\ge$T9 dwarf discoveries within this volume is still incomplete.}
\end{deluxetable}

\clearpage

\clearpage

\begin{figure}
\epsscale{0.8}
\figurenum{2}
\plotone{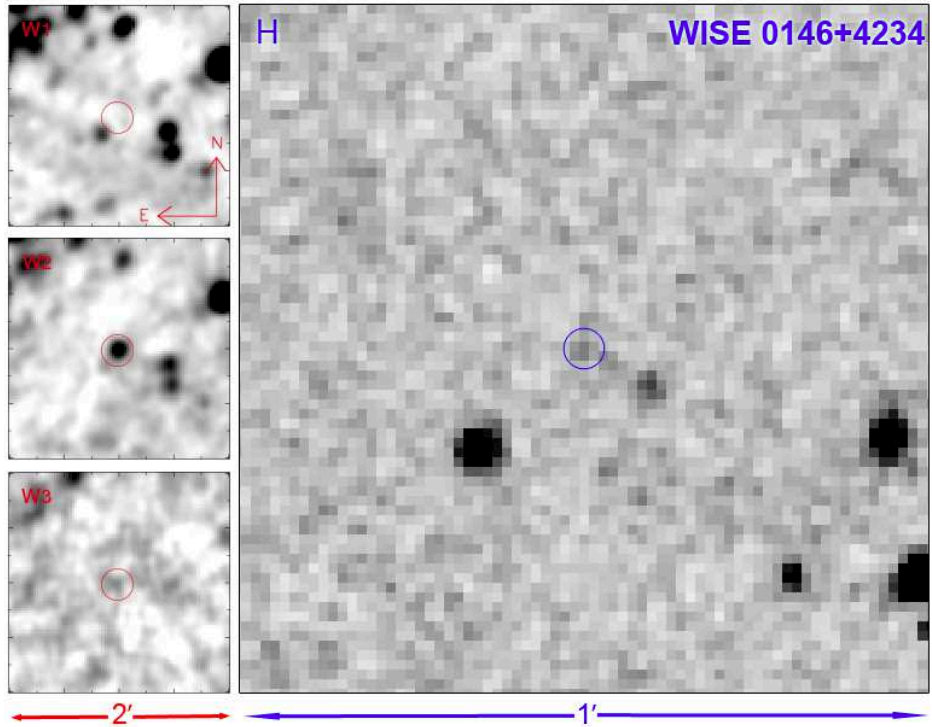}
\caption{Finder charts for all thirteen known WISE Y dwarf discoveries. Along the left are 2$\times$2 arcminute images from WISE in bands W1, W2, and W3 with the location of the Y dwarf shown by the red circle. The large image on the right is a 1$\times$1 arcminute zoom taken in the the near-infrared; the Y dwarf is circled in blue. In all images, north is up and east is to the left. Note that the field for WISE 0535$-$7500 is located toward the outskirts of the Large Magellanic Cloud, which is the reason for the high source density. For WISE 0713$-$2917, our near-infrared image on the right does not completely fill the 1$\times$1 arcminute field allotted for it. For WISE 1541$-$2250, the prominent white ring on the large image is a latent artifact caused by a bright star during the dithering process.
\label{Finder0146+4234}}
\end{figure}

\clearpage

\begin{figure}
\epsscale{0.8}
\figurenum{2}
\plotone{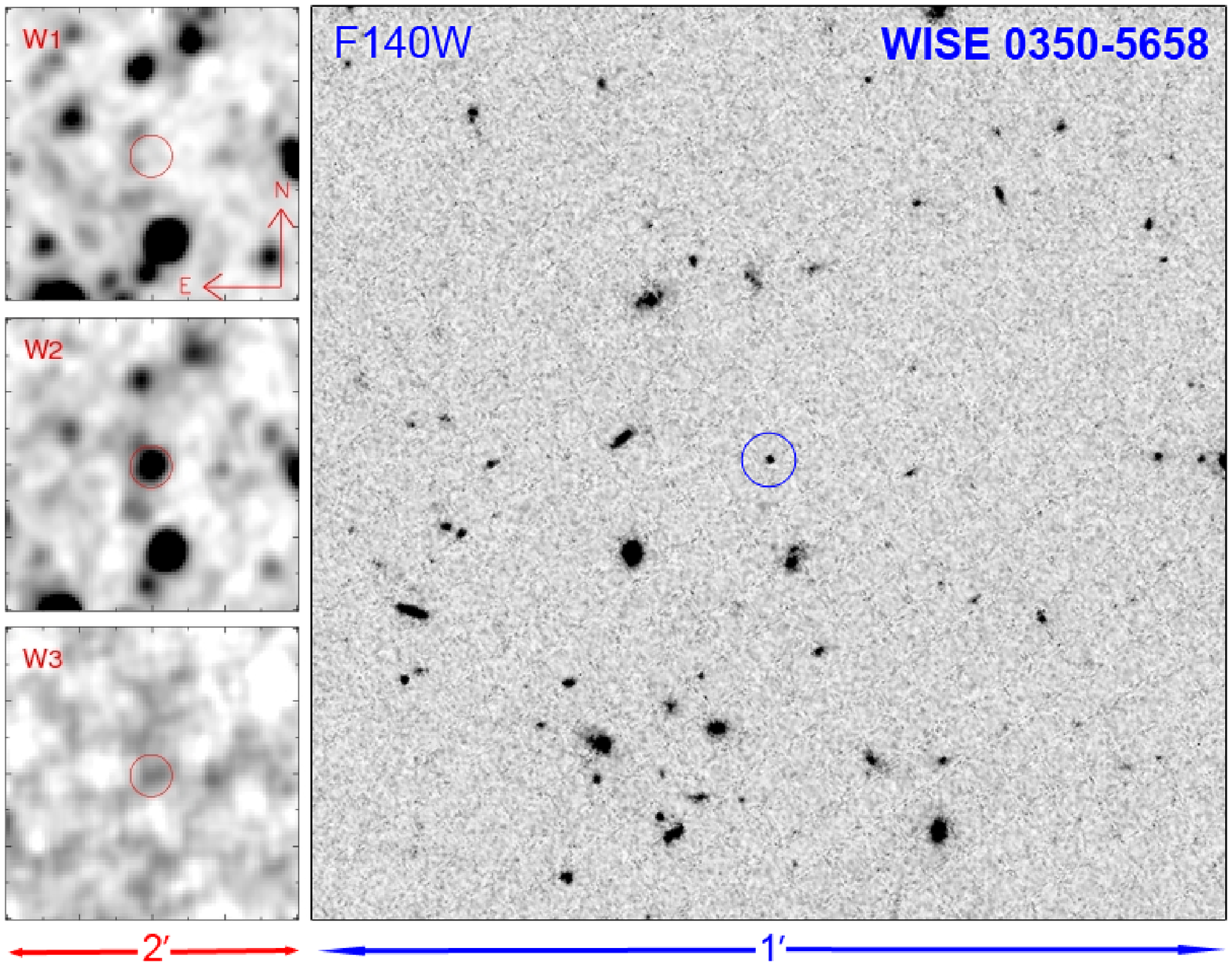}
\caption{Continued.
\label{Finder0350-5658}}
\end{figure}

\clearpage

\begin{figure}
\epsscale{0.8}
\figurenum{2}
\plotone{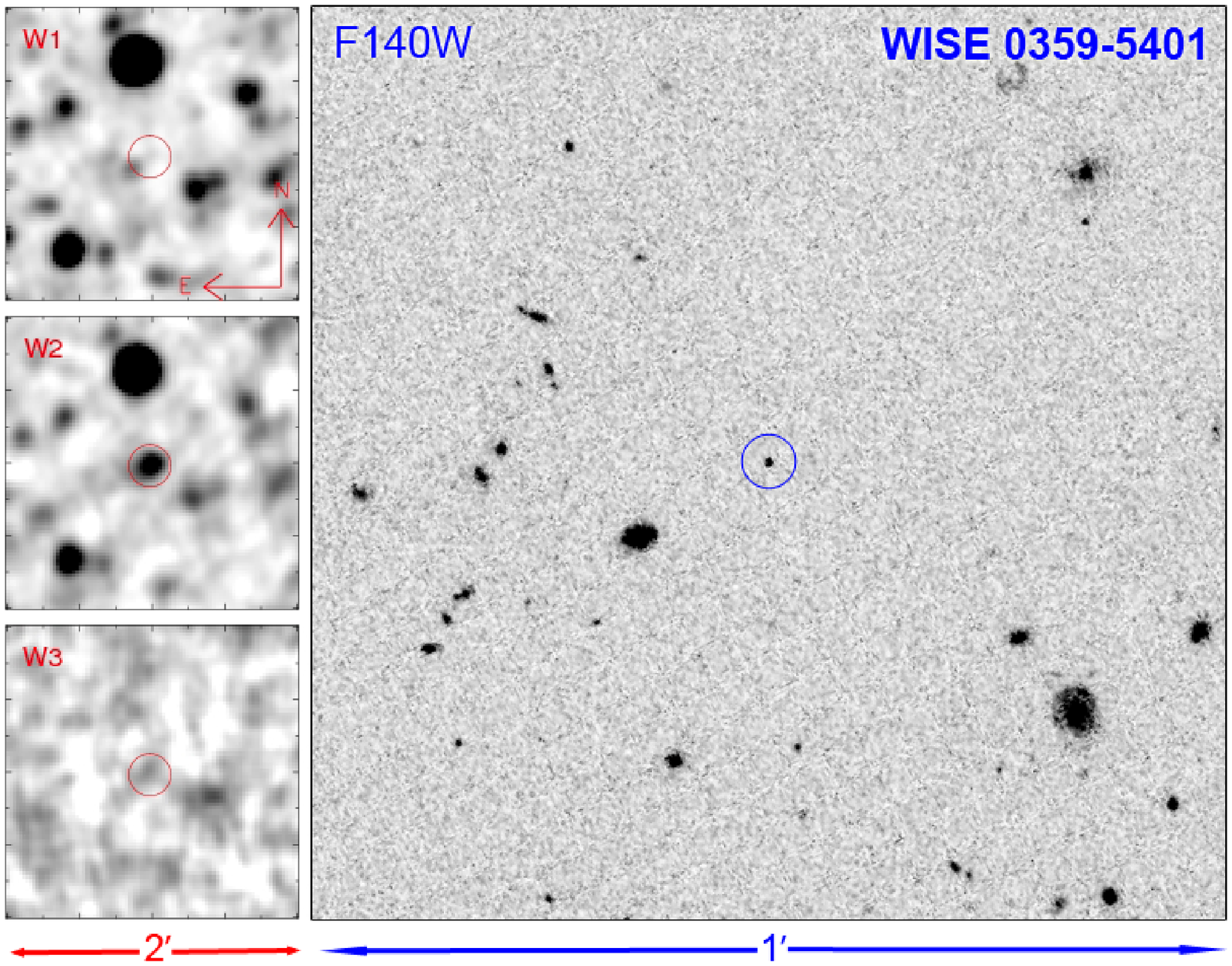}
\caption{Continued.
\label{Finder0359-5401}}
\end{figure}

\clearpage

\begin{figure}
\epsscale{0.8}
\figurenum{2}
\plotone{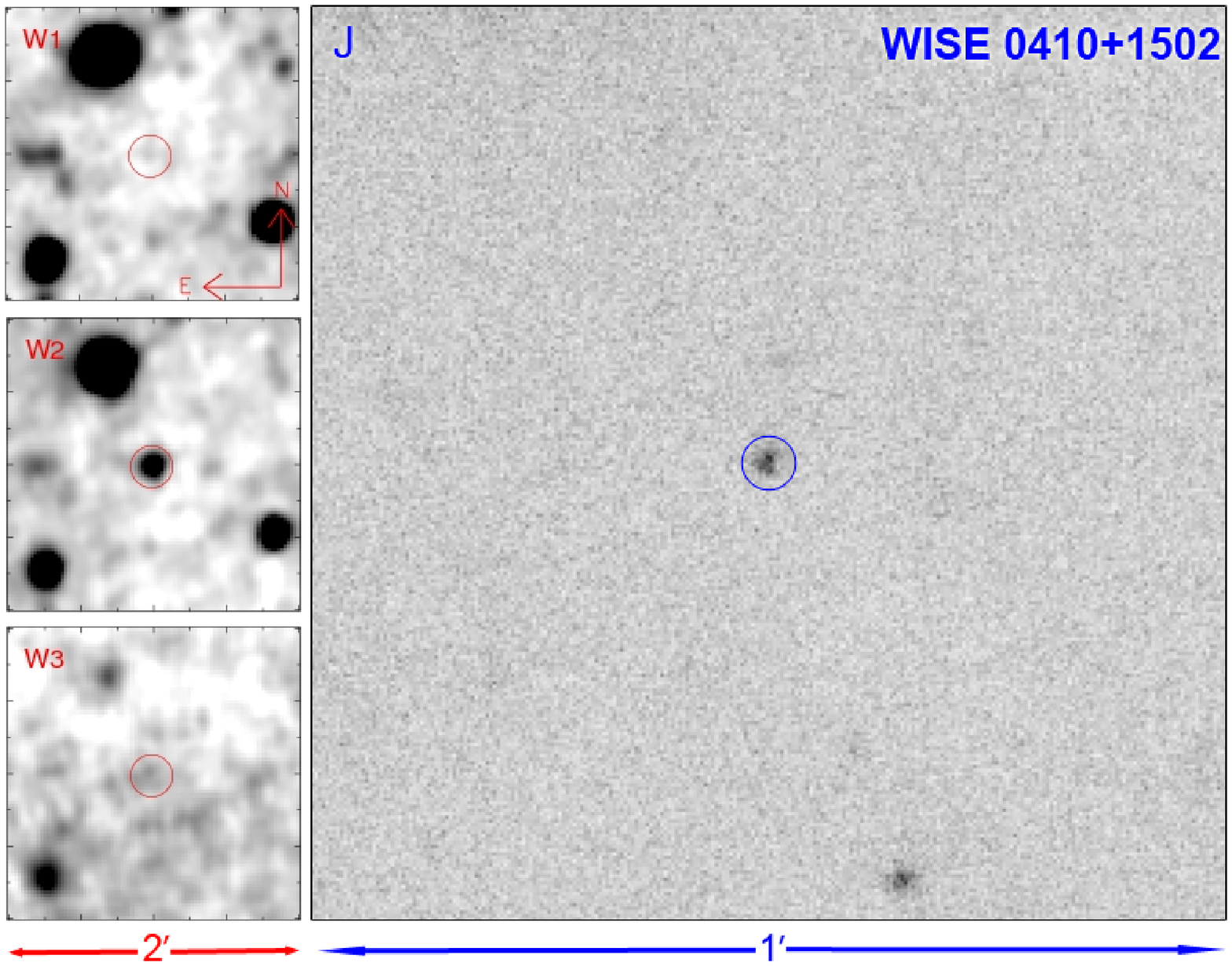}
\caption{Continued.
\label{Finder0410+1502}}
\end{figure}

\clearpage

\begin{figure}
\epsscale{0.8}
\figurenum{2}
\plotone{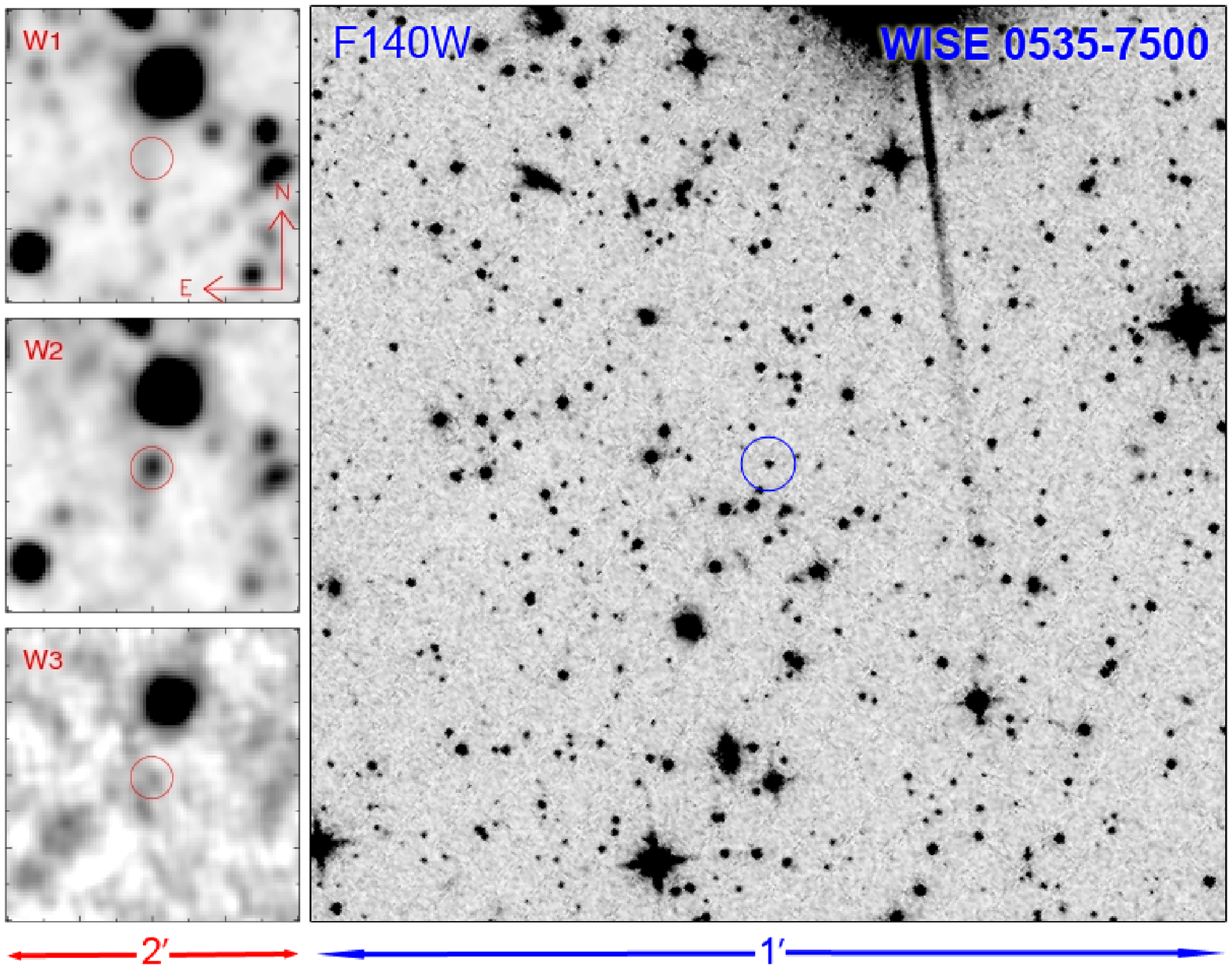}
\caption{Continued.
\label{Finder0535-7500}}
\end{figure}

\clearpage

\begin{figure}
\epsscale{0.8}
\figurenum{2}
\plotone{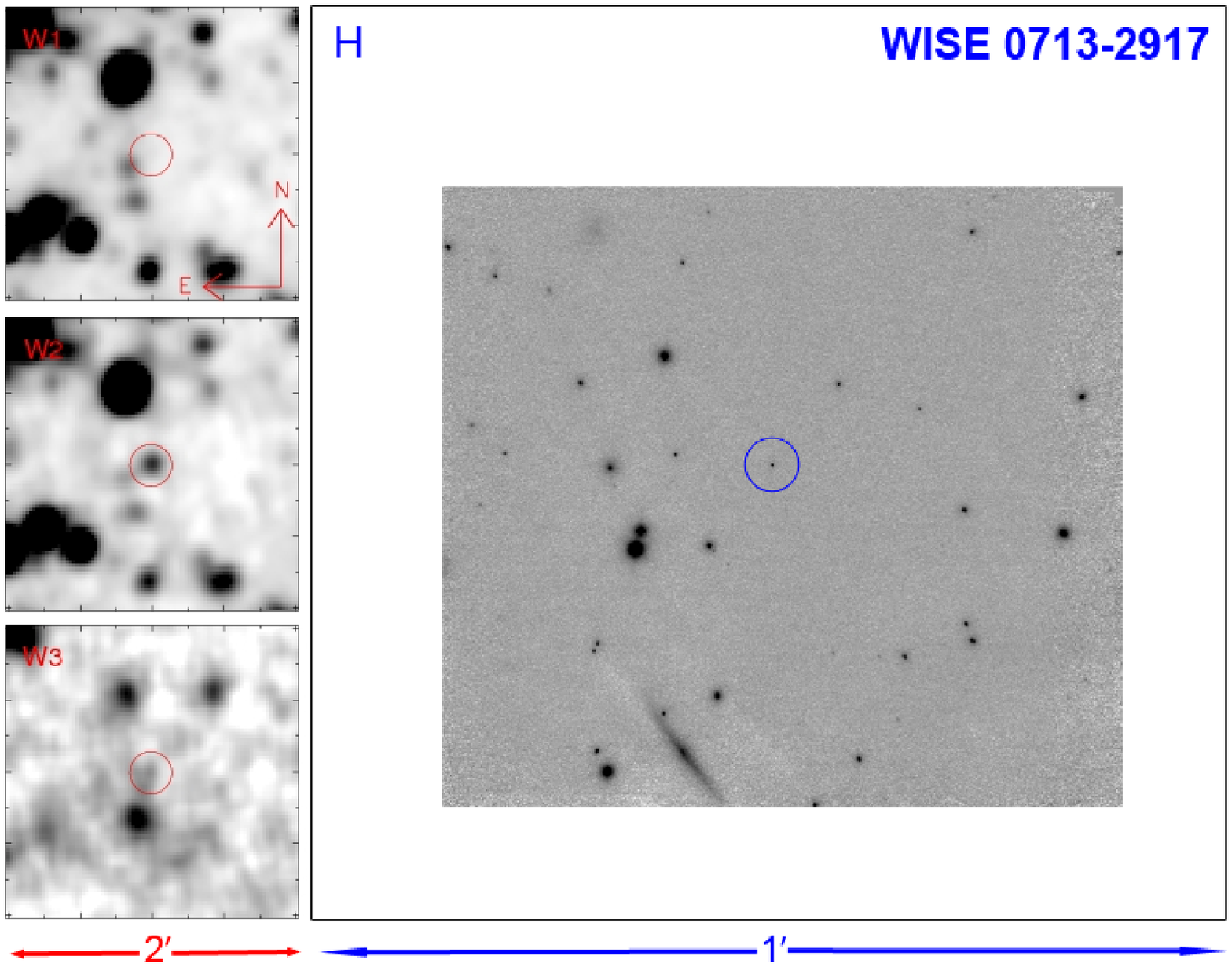}
\caption{Continued. 
\label{Finder0713-2917}}
\end{figure}

\clearpage

\begin{figure}
\epsscale{0.8}
\figurenum{2}
\plotone{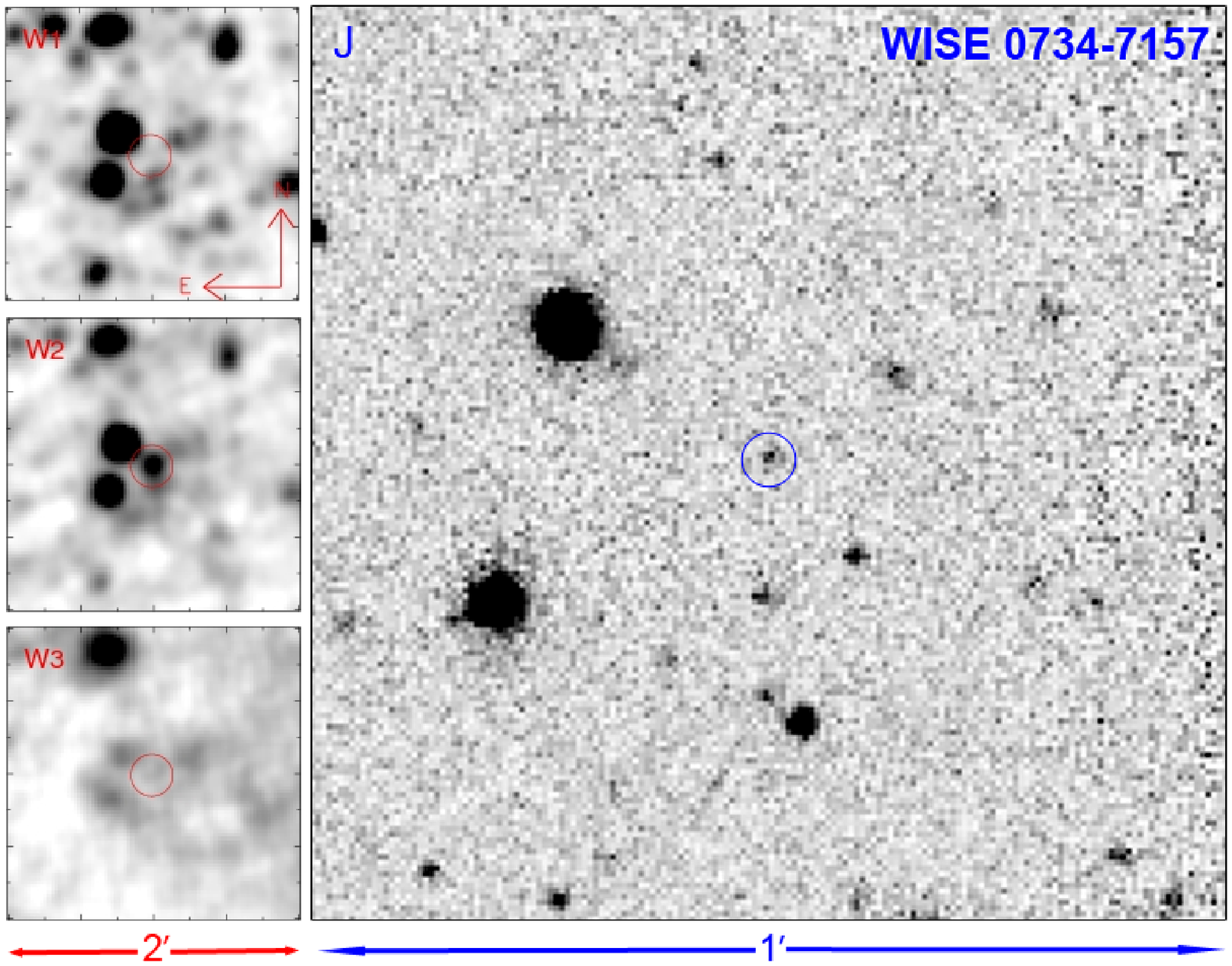}
\caption{Continued.
\label{Finder0734-7157}}
\end{figure}

\clearpage

\begin{figure}
\epsscale{0.8}
\figurenum{2}
\plotone{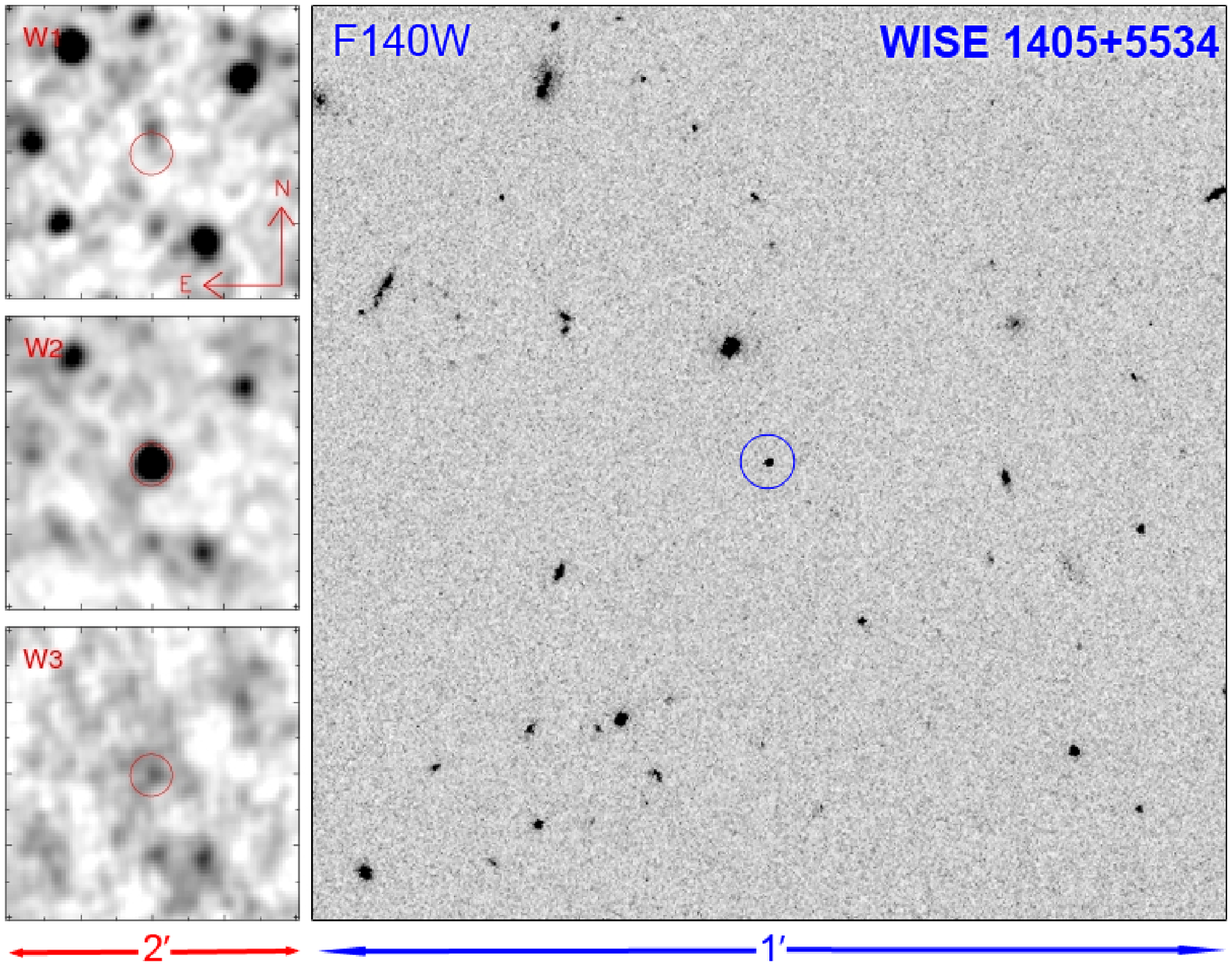}
\caption{Continued.
\label{Finder1405+5534}}
\end{figure}

\clearpage

\begin{figure}
\epsscale{0.8}
\figurenum{2}
\plotone{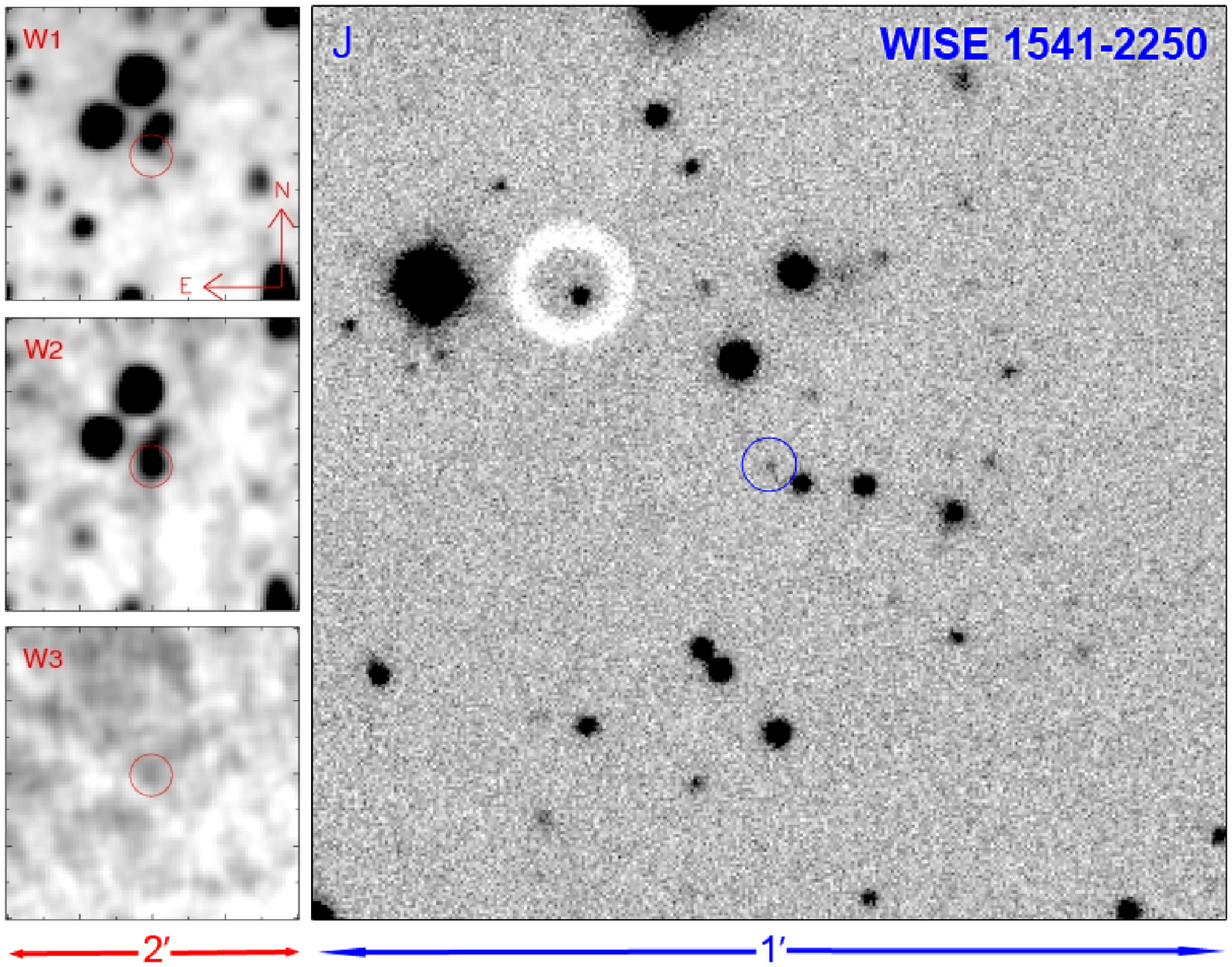}
\caption{Continued.
\label{Finder1541-2250}}
\end{figure}

\clearpage

\begin{figure}
\epsscale{0.8}
\figurenum{2}
\plotone{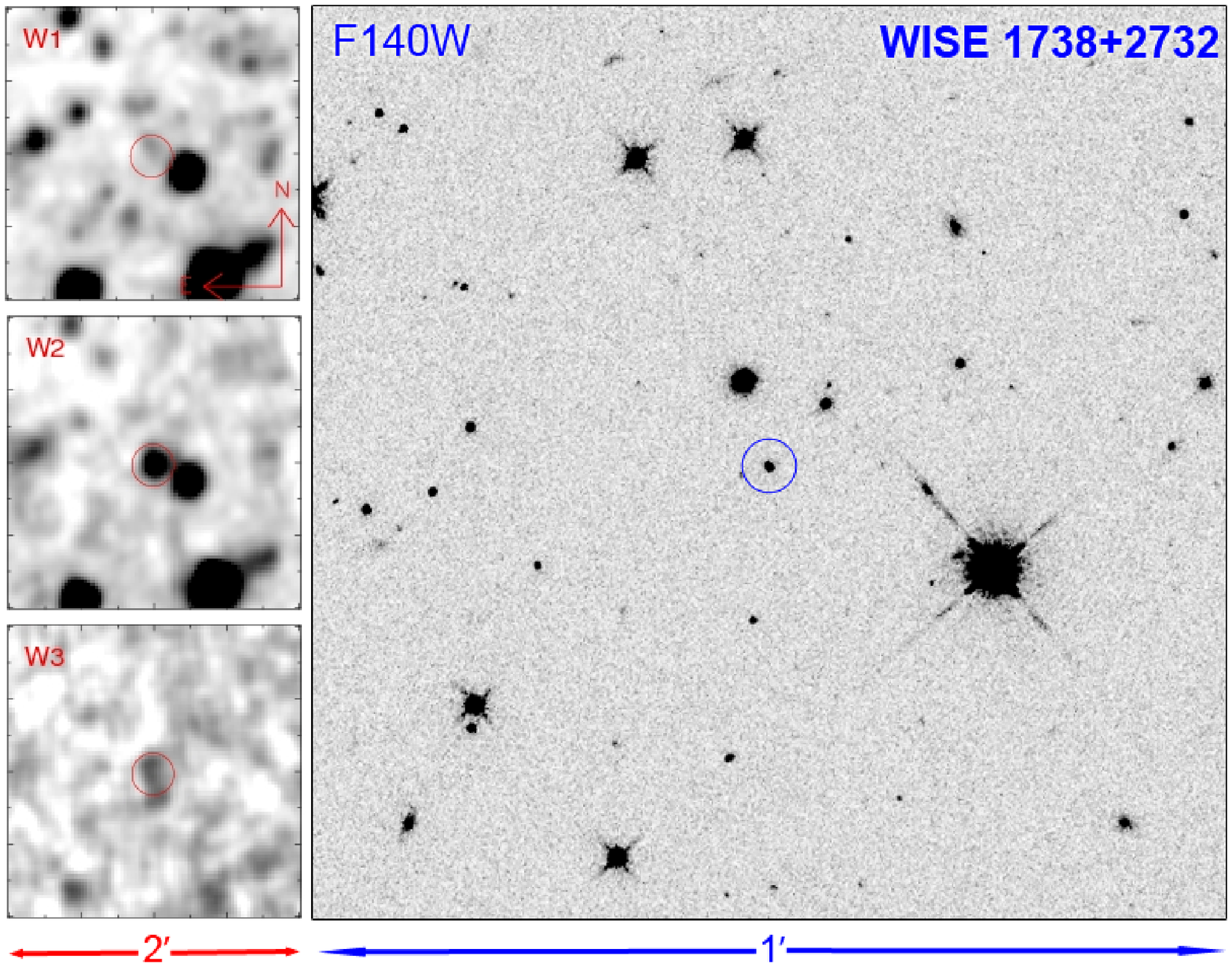}
\caption{Continued.
\label{Finder1738+2732}}
\end{figure}

\clearpage

\begin{figure}
\epsscale{0.8}
\figurenum{2}
\plotone{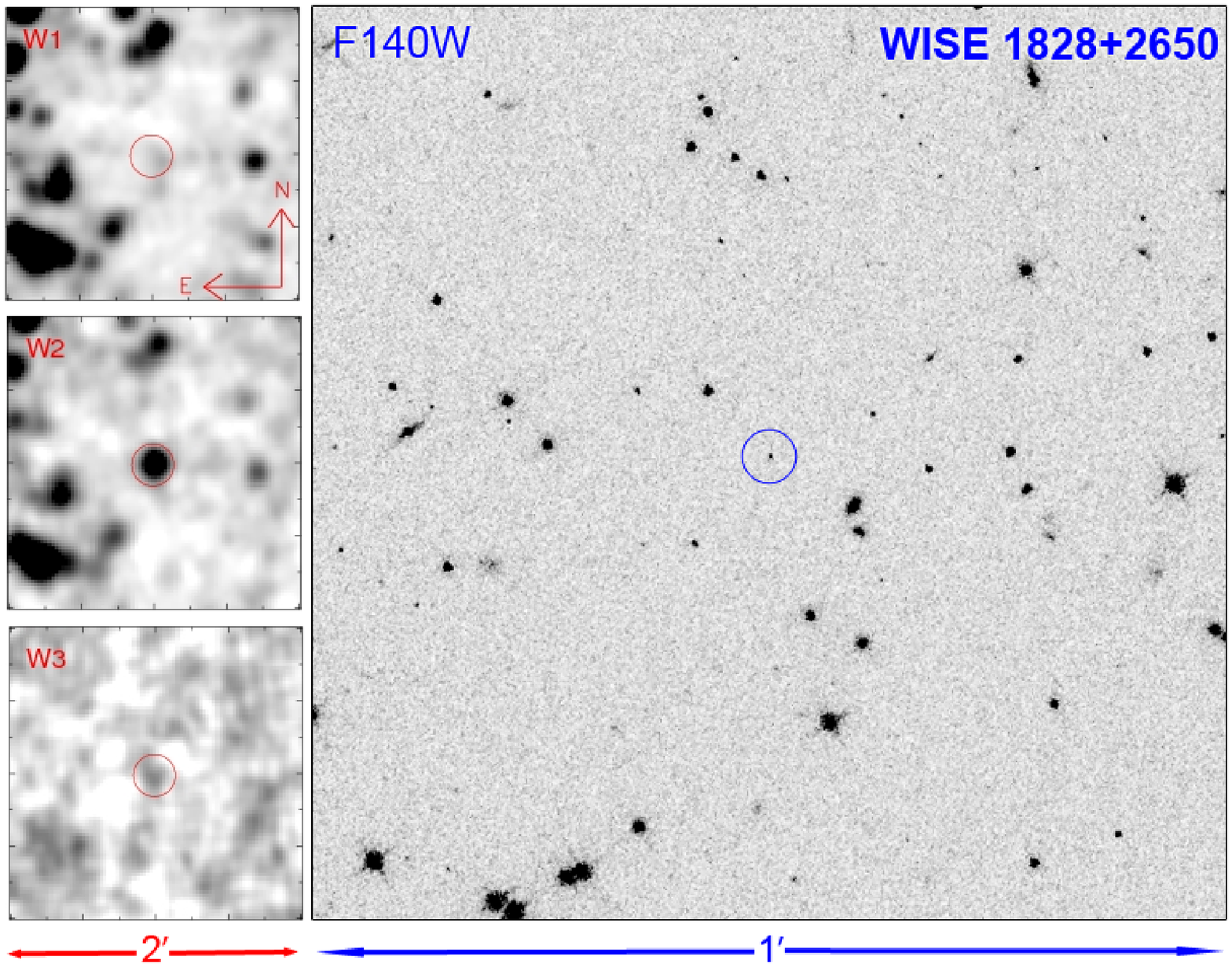}
\caption{Continued.
\label{Finder1828+2650}}
\end{figure}

\clearpage

\begin{figure}
\epsscale{0.8}
\figurenum{2}
\plotone{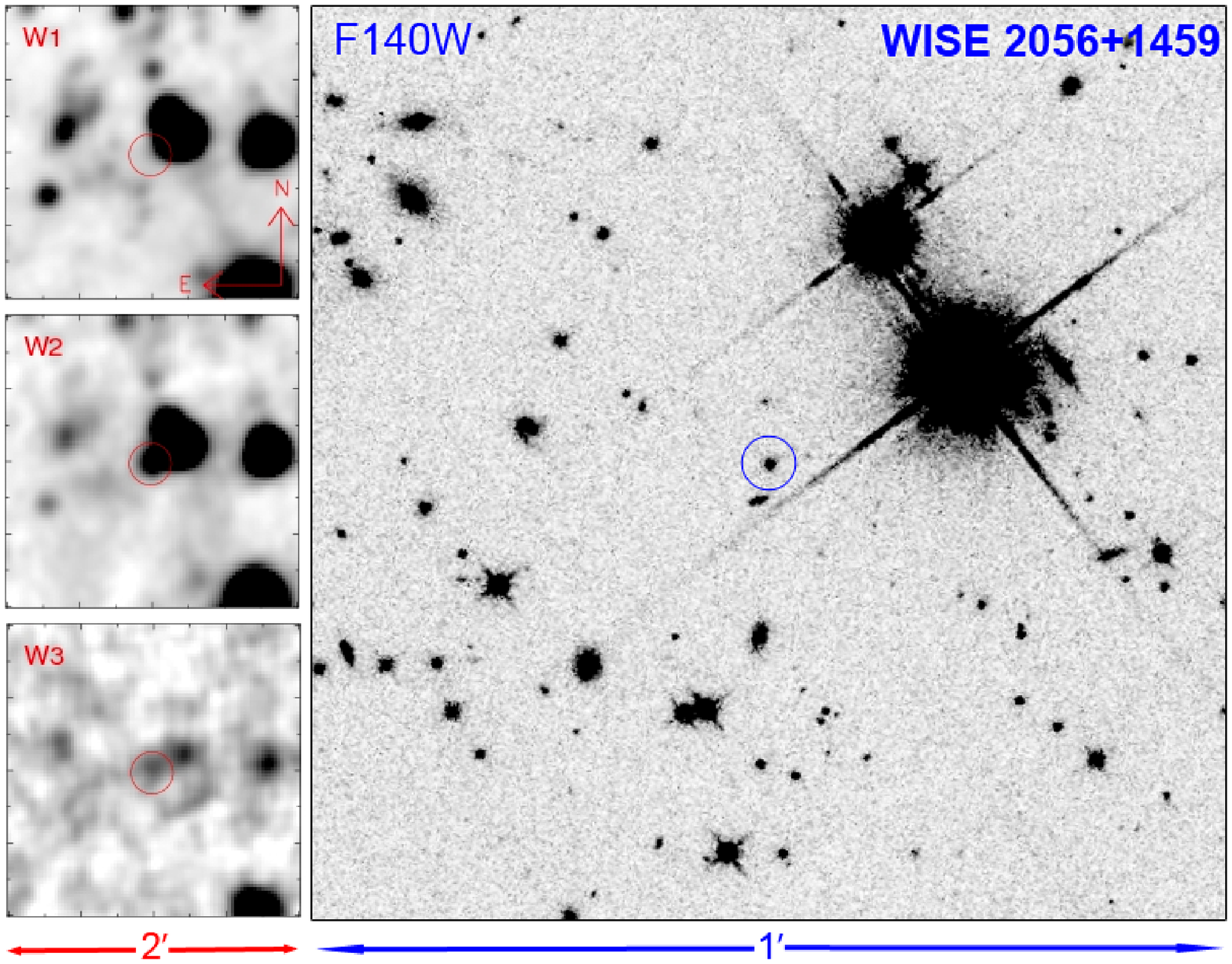}
\caption{Continued.
\label{Finder2056+1459}}
\end{figure}

\clearpage

\begin{figure}
\epsscale{0.8}
\figurenum{2}
\plotone{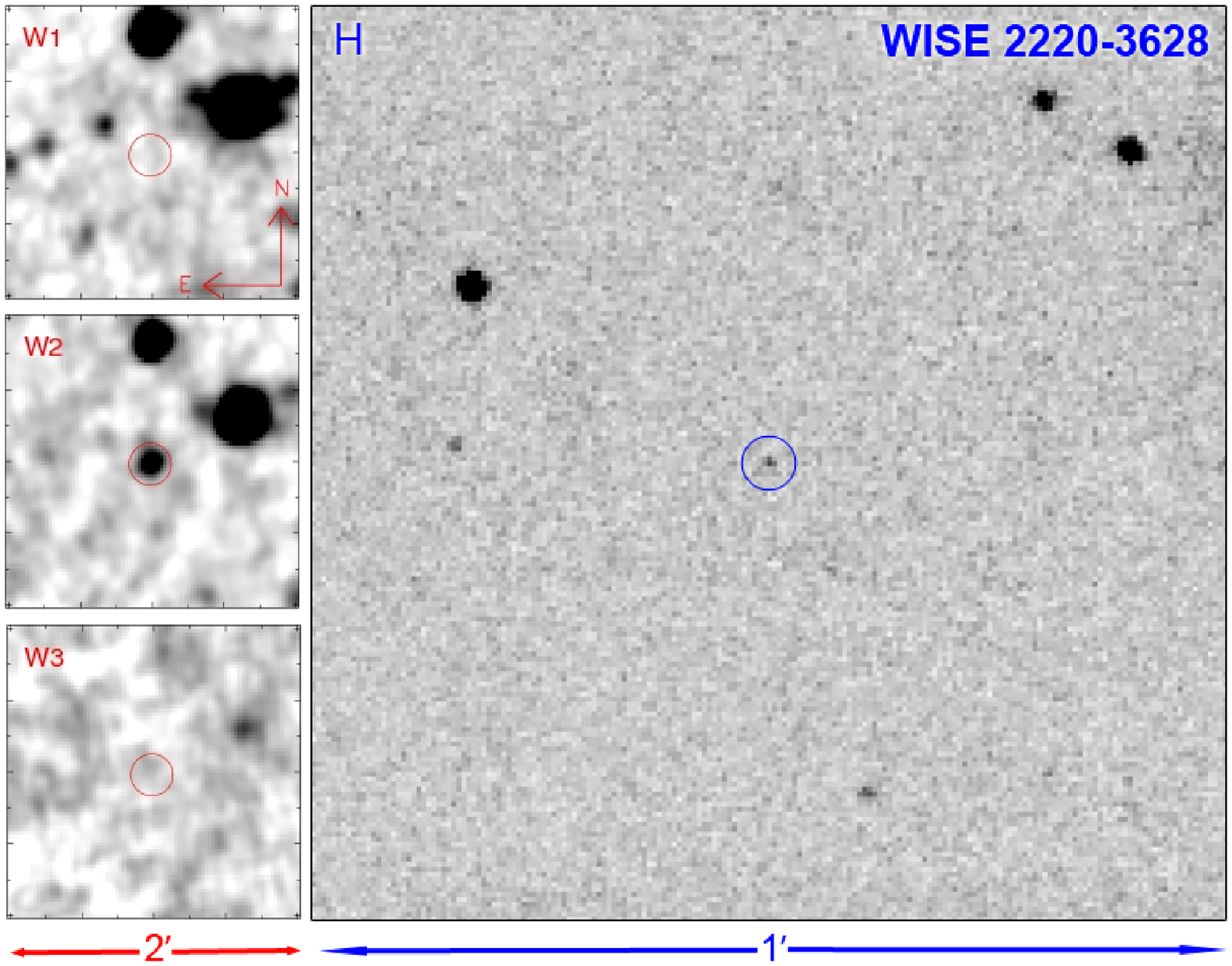}
\caption{Continued.
\label{Finder2220-3628}}
\end{figure}

\clearpage

\begin{figure}
\epsscale{0.8}
\figurenum{3}
\plotone{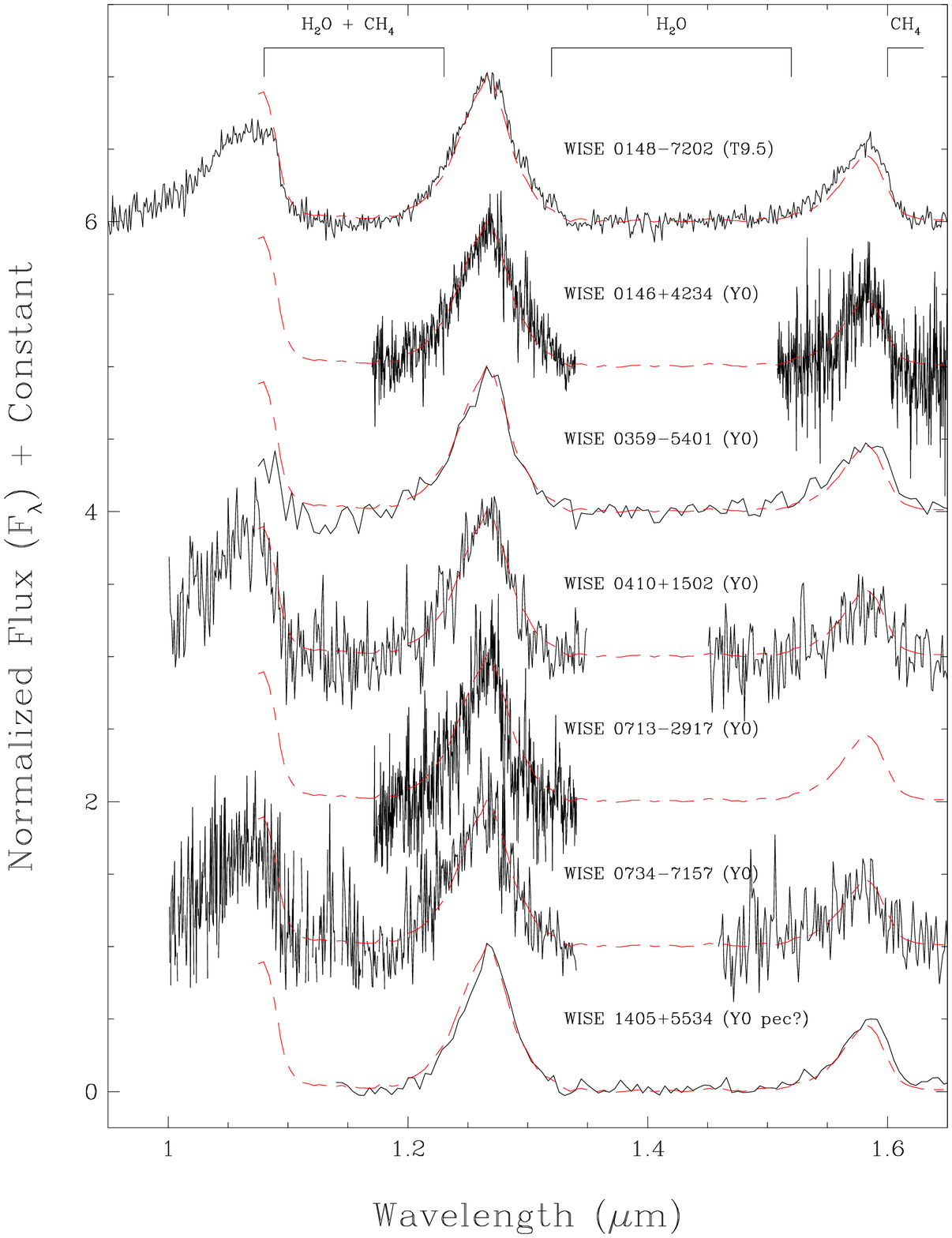}
\caption{Along with Fig~\ref{YDwarfs2} and Fig~\ref{YDwarfs3}, the spectra of all known Y dwarfs together with a comparison
spectrum of the T9.5 dwarf WISE 1048-7202. Each spectrum is normalized to one at its peak in the $J$-band and integral 
offsets have been added to separate the spectra vertically. Overplotted on each spectrum is the Y0 dwarf spectral standard
WISE 1738+2732 (dashed red curve). For NIRSPEC data taken in the N5 configuration ($H$-band), the normalization has been set 
so that the $H$-band peak of the Y dwarf matches the $H$-band peak of the spectral standard.
\label{YDwarfs1}}
\end{figure}

\clearpage

\begin{figure}
\epsscale{0.8}
\figurenum{4}
\plotone{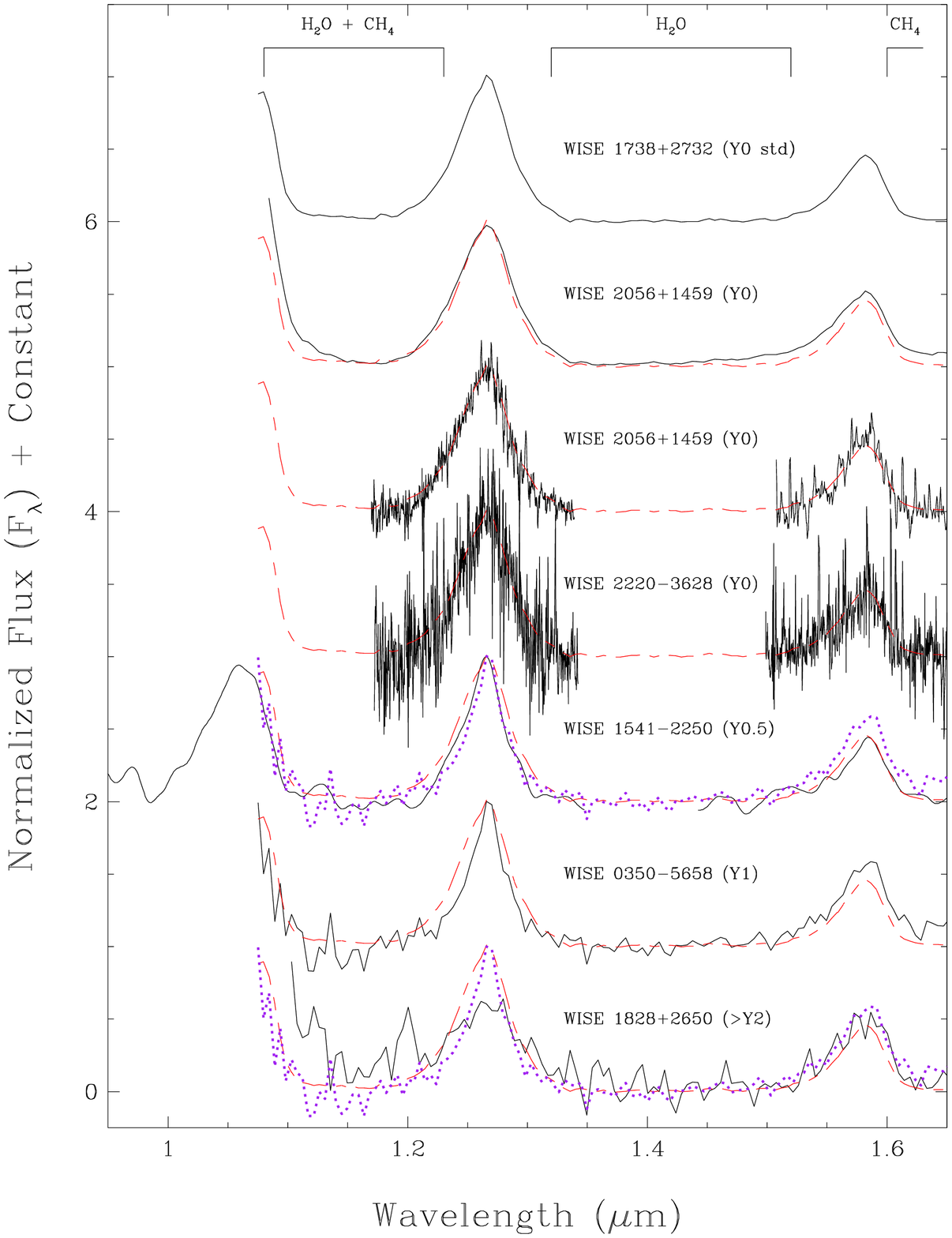}
\caption{Spectra of Y dwarfs (continued). The spectrum of the Y1 standard, WISE 0350-5658, is shown by the purple dotted line and is normalized to one at the $J$-band peak. See caption to Fig~\ref{YDwarfs1} for other details.
\label{YDwarfs2}}
\end{figure}

\clearpage

\begin{figure}
\epsscale{0.8}
\figurenum{5}
\plotone{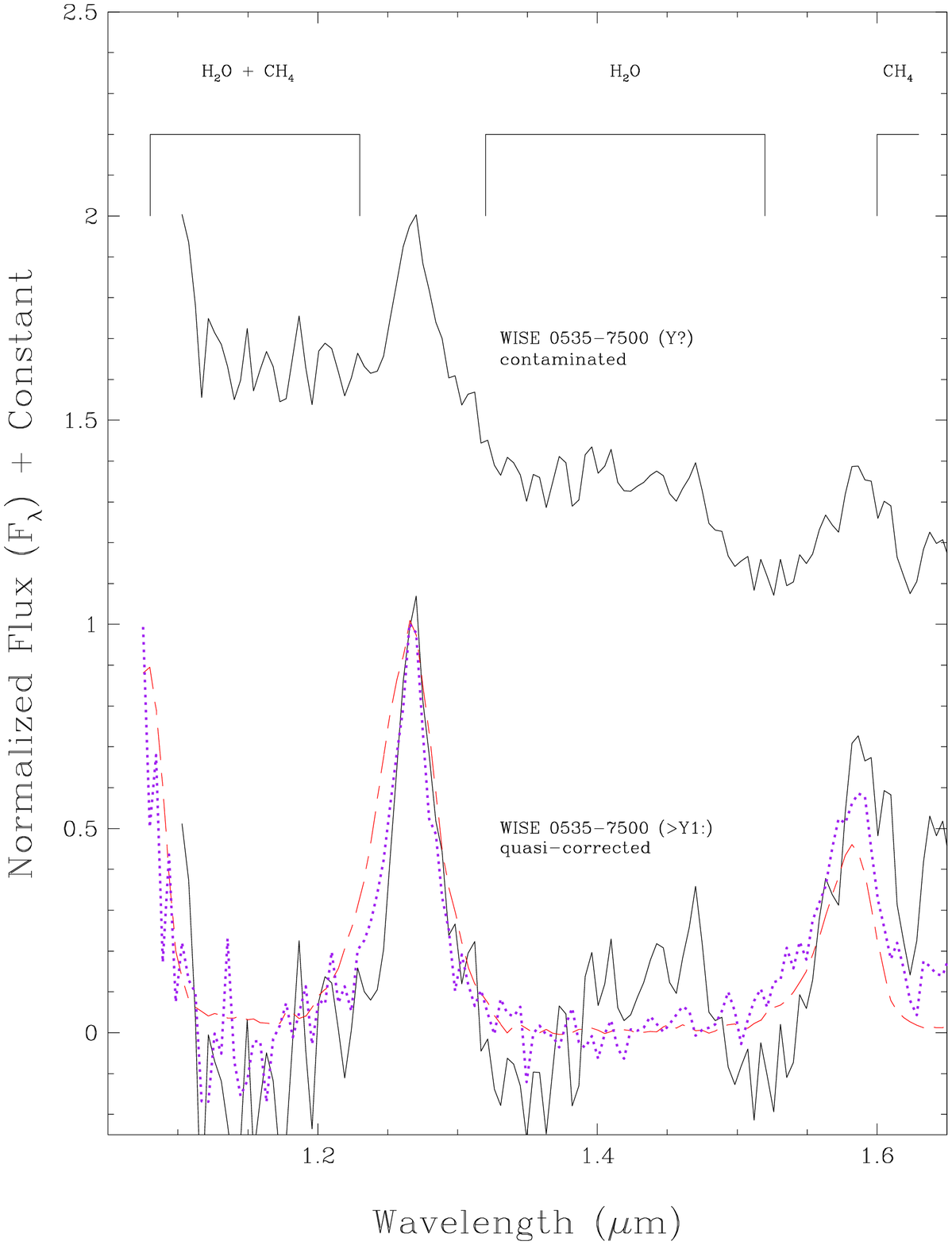}
\caption{Spectra of Y dwarfs (continued). The spectrum of WISE 0535-7500 is shown in black both before (above) and after (below) subtraction of the overlapping field spectrum. Both spectra are normalized to one at the peak of $J$-band, and the top spectrum is offset by one in the vertical direction to separate it from the lower spectrum. See caption to Fig~\ref{YDwarfs1} and Fig~\ref{YDwarfs2} for other details, and see the text for a description of the correction applied.
\label{YDwarfs3}}
\end{figure}

\clearpage

\begin{figure}
\epsscale{0.8}
\figurenum{6}
\plotone{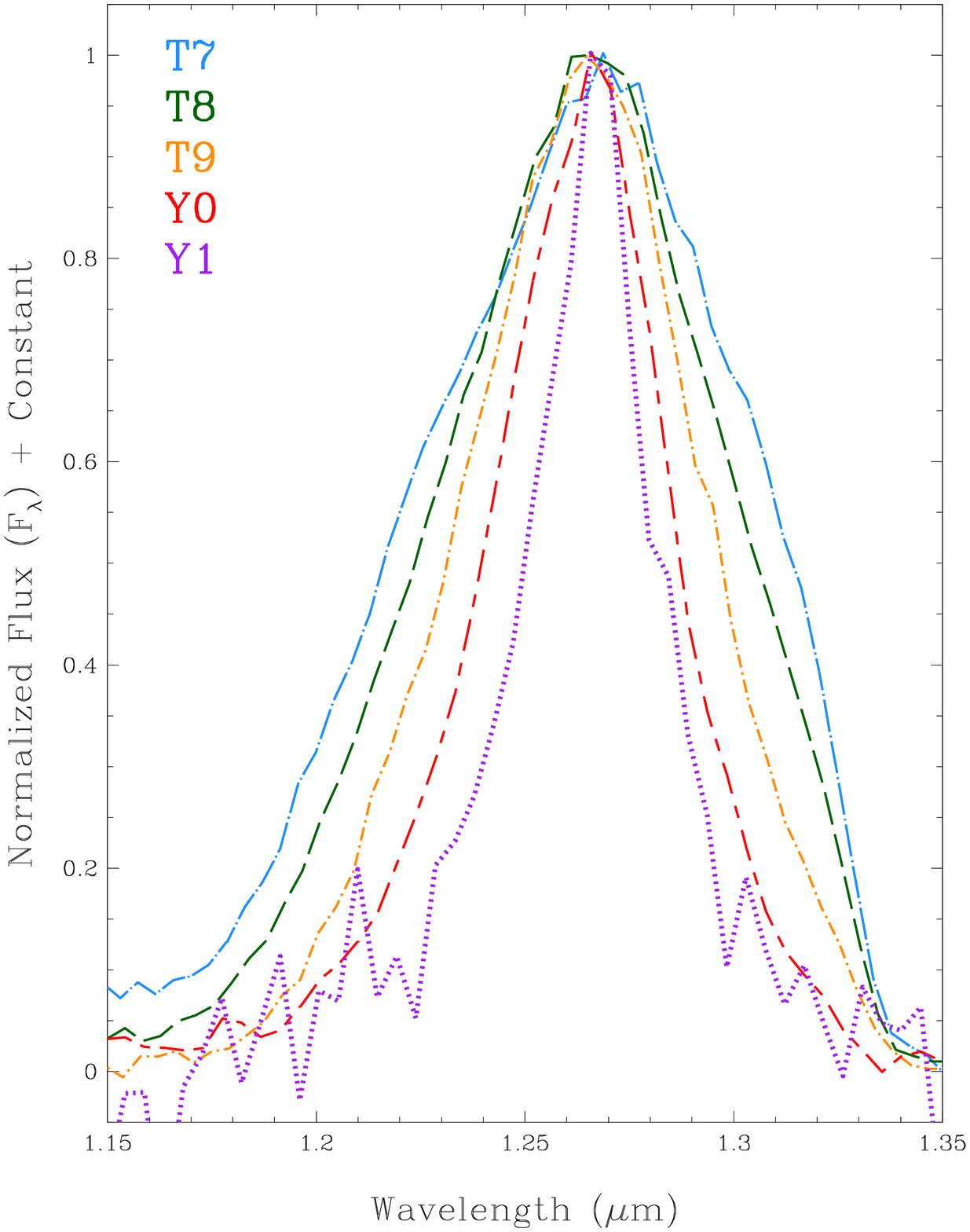}
\caption{Overplots of the $J$-band spectra of the T7, T8, T9, and Y0 standards 2MASS 0727+1710 (light blue, dot/long-dash; \citealt{burgasser2002}), 2MASS 0415$-$0935 (green, dash; \citealt{burgasser2002}), UGPS 0722$-$0540 (gold, dot/short-dash; \citealt{lucas2010}), and WISE 1732+2732 (red, long-dash/short-dash; \citealt{cushing2011}), respectively, along with our proposed Y1 standard, WISE 0350$-$5658 (purple, dot). All spectra are normalized to one at their peak flux.
\label{YDwarfs_Jcomparison}}
\end{figure}

\clearpage

\begin{figure}
\includegraphics[scale=0.65,angle=270]{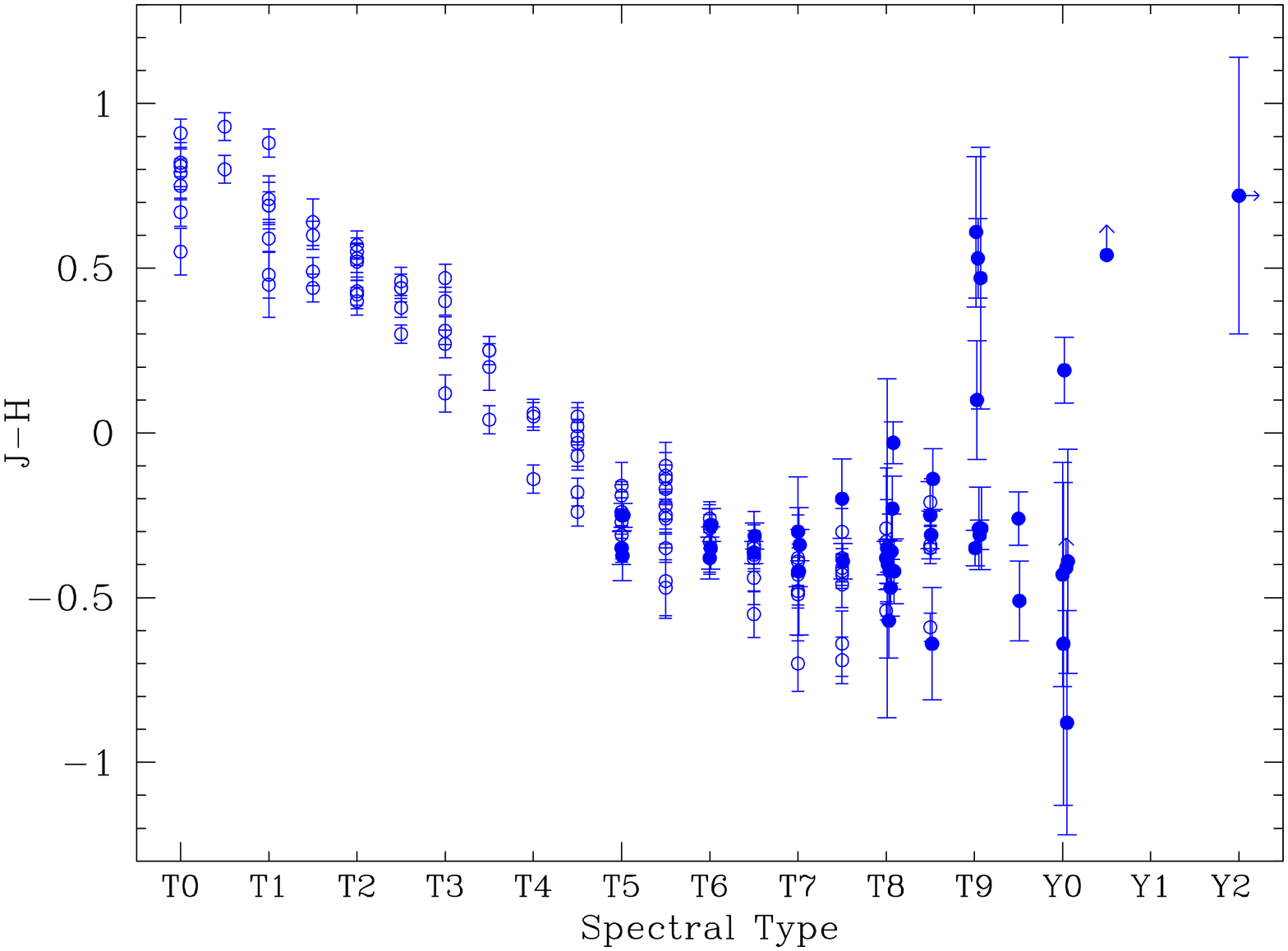}
\figurenum{7}
\caption{A plot of the $J-H$ color (on the MKO-NIR filter system) as a function of spectral type from T0 through Y2. Small offsets have been added to some of the spectral type values to ease visualization of individual data points. Colors for Y dwarfs from Table~\ref{y_dwarf_discoveries_phot} and T dwarfs from \cite{kirkpatrick2011} and Mace et al.\ (in prep.) are shown by solid blue points. Colors of other T dwarfs from \cite{leggett2010} are shown by open blue points.
\label{YDwarfs_Colors2}}
\end{figure}

\clearpage

\begin{figure}
\includegraphics[scale=0.65,angle=270]{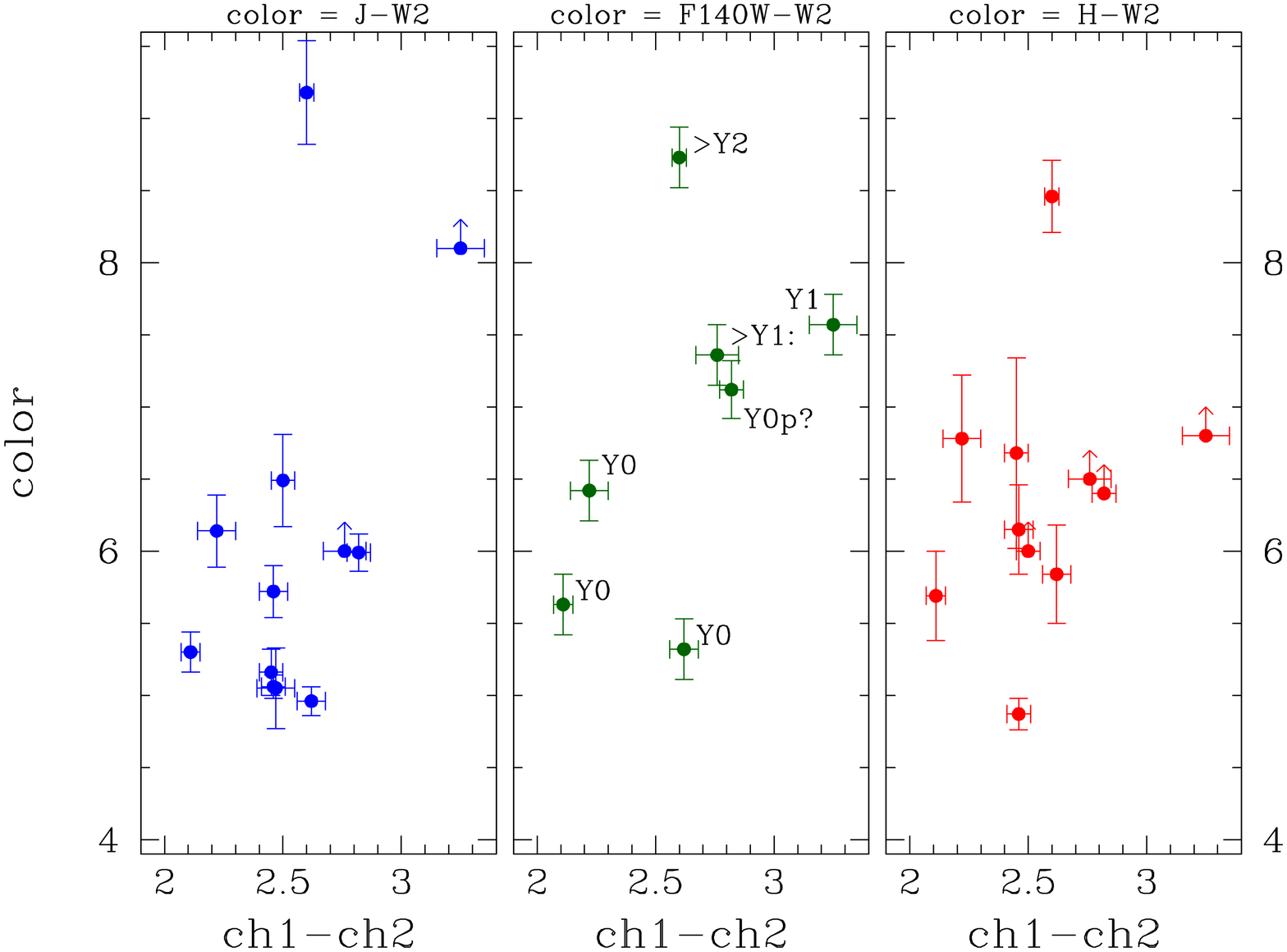}
\figurenum{8}
\caption{Various near-infrared/WISE colors plotted as a function of the {\it Spitzer} ch1$-$ch2 color. From left to right are shown the $J-$W2 color (blue), F140W$-$W2 color (green), and $H-$W2 color (red). All panels are shown with the same scales on the $x$ and $y$ axes. In the F140W$-$W2 panel, spectral types are shown next to each of the points.
\label{YDwarfs_Colors}}
\end{figure}

\clearpage

\begin{figure}
\includegraphics[scale=0.65,angle=270]{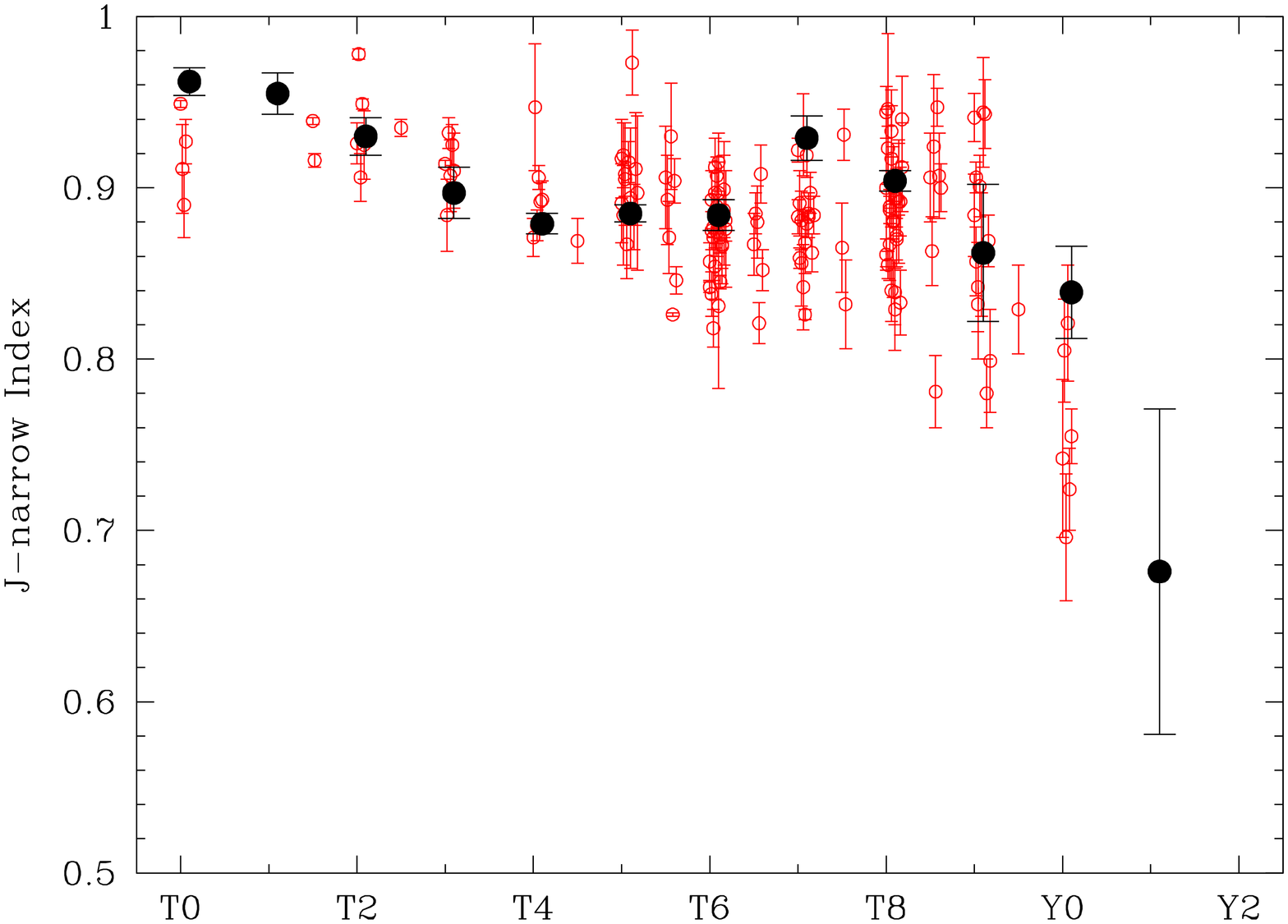}
\figurenum{9}
\caption{Values of our newly defined J-narrow index as a function of spectral type. Objects are taken from this paper, Mace et al.\ (in prep.), \cite{cushing2011}, and \cite{kirkpatrick2011}. See Mace et al.\ for details concerning these computations. Spectral standards are shown by solid black points and other objects are shown by open red points. Small toggles have been added to the spectral type values to reduce the amount of overlap in each spectral bin. With the exception of the Y1 standard, only those objects with $J$-narrow errors of less than 0.05 are plotted.
\label{YDwarfs_Indices}}

\end{figure}

\clearpage

\begin{figure}
\epsscale{0.8}
\figurenum{10}
\plotone{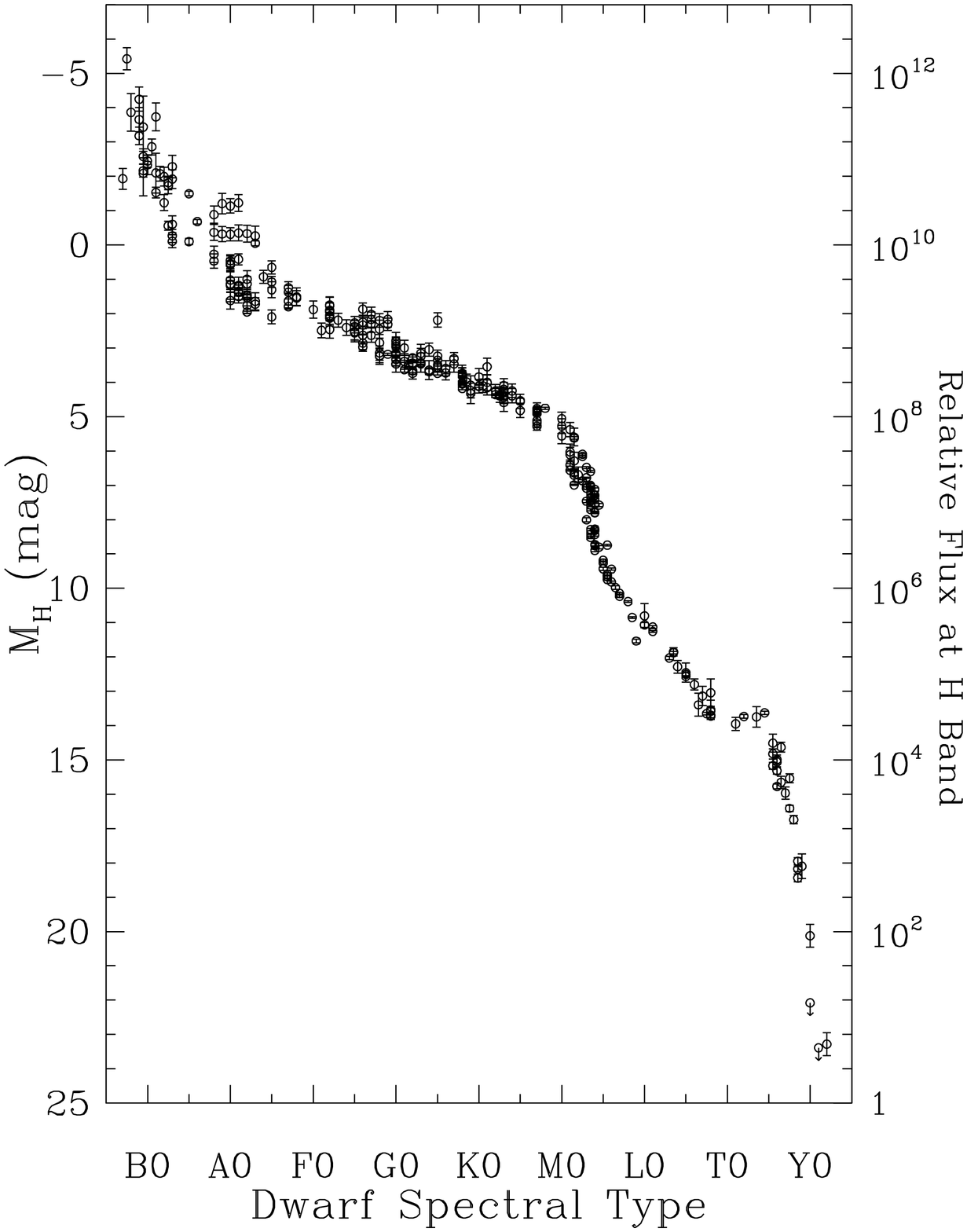}
\caption{A Hertzsprung-Russell diagram at $H$-band showing the Y dwarfs in context with other field brown dwarfs and main sequence stars. The intrinsically faintest Y dwarf so far recognized is roughly twelve orders of magnitude fainter than an O-type main sequence star at this wavelength. See Section 4.1 for details about the sample plotted.
\label{MH_type}}
\end{figure}

\clearpage

\begin{figure}
\includegraphics[scale=0.65,angle=270]{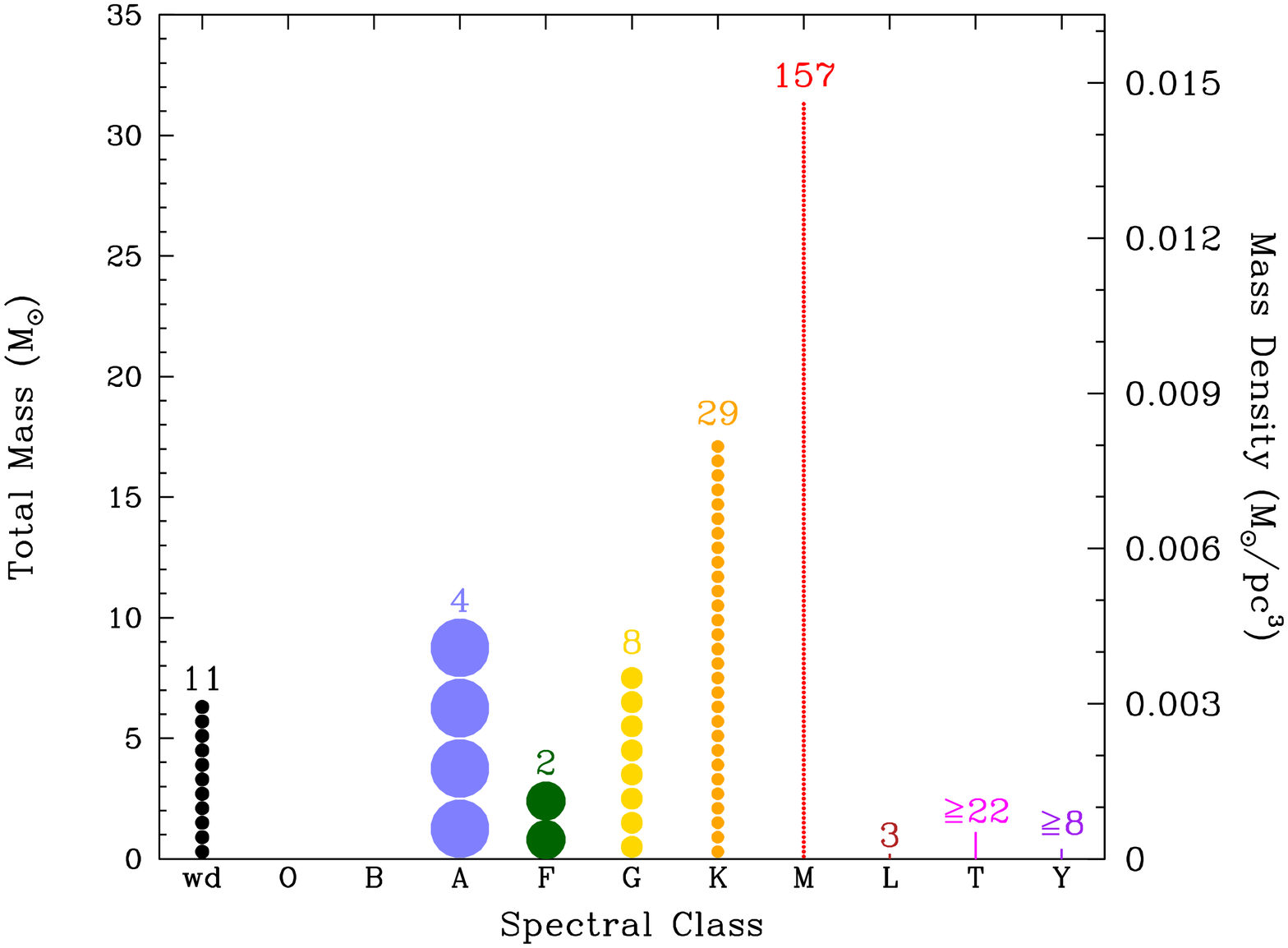}
\figurenum{11}
\caption{The 8pc sample as a function of spectral type plotted in three ways -- as total mass (left axis), mass density (right axis), and histogram (numbers above each star stack). White dwarfs are shown in black, A stars in light blue, F stars in green, G stars in yellow, K stars in orange, M dwarfs in red, L dwarfs in firebrick, T dwarfs in magenta, and Y dwarfs in purple. The only bins believed to suffer from significant incompleteness are those of the T and Y dwarfs; it is likely that a small number of solivagant T dwarfs, and a larger number of Y dwarfs, have yet to be identified along with T and Y companions to higher-mass objects already known within the 8 pc volume.
\label{8pc_sample}}
\end{figure}
\clearpage

\begin{figure}
\epsscale{0.8}
\figurenum{12}
\plotone{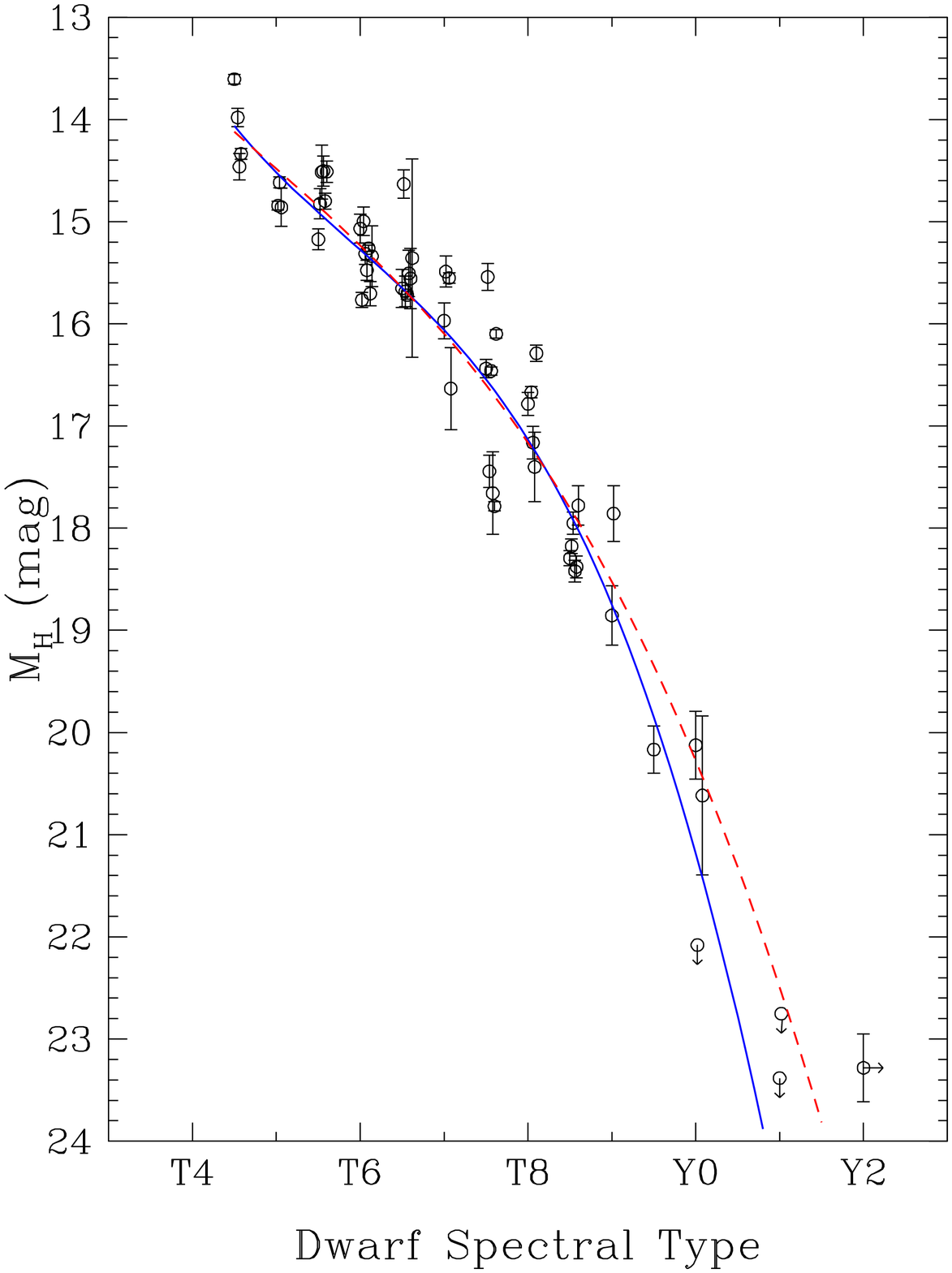}
\caption{A plot of the absolute $H$-band magnitude of dwarfs with spectral types $>$T4 and measured trigonometric parallaxes, from Table~\ref{TYparallaxes}. Third-order fits to the data are shown by the dashed red curve (which includes the point for WISE 1828+2650 at lower right) and the solid blue curve (which excludes WISE 1828+2650). See Section 4.3 for details.
\label{MH_TYtype}}
\end{figure}

\clearpage

\begin{figure}
\epsscale{0.8}
\figurenum{13}
\plotone{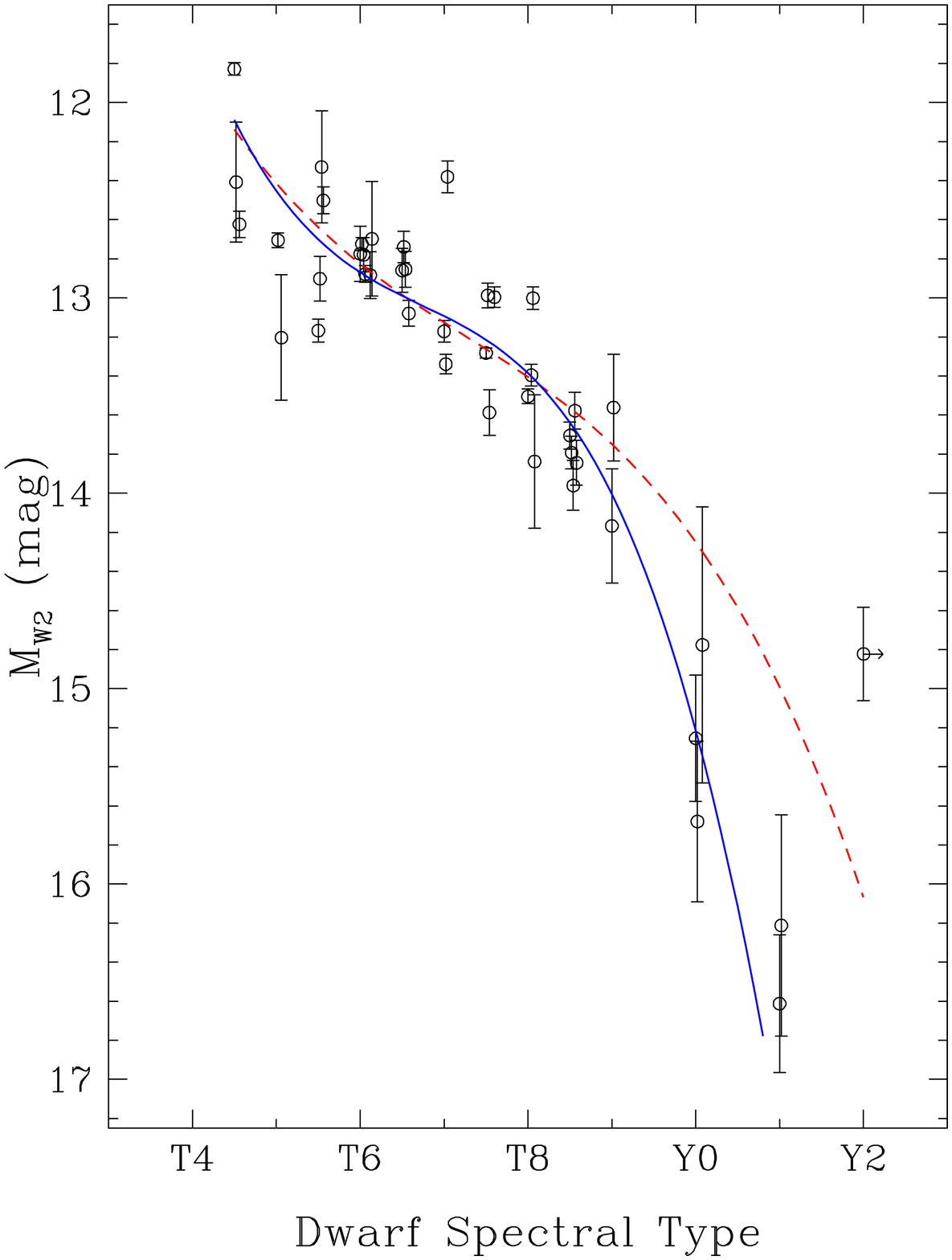}
\caption{A plot of the absolute W2-band magnitude of dwarfs with spectral types $>$T4 and measured trigonometric parallaxes, from Table~\ref{TYparallaxes}. Third-order fits to the data are shown by the dashed red curve (which includes the point for WISE 1828+2650 near the right edge) and the solid blue curve (which excludes WISE 1828+2650). See Section 4.3 for details.
\label{MW2_TYtype}}
\end{figure}

\clearpage

\begin{figure}
\epsscale{0.8}
\figurenum{14}
\plotone{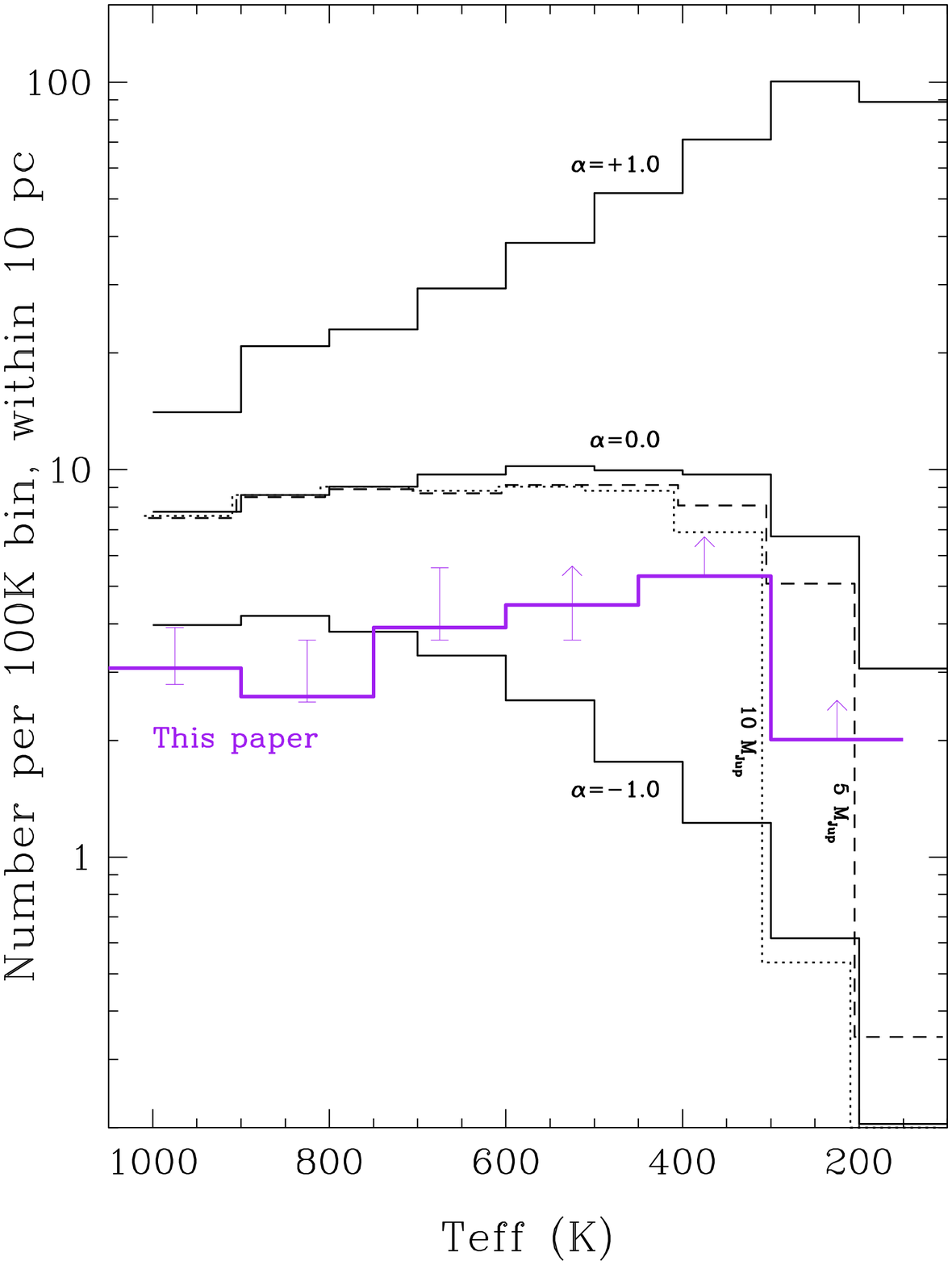}
\caption{The predicted number of brown dwarfs within 10 pc for three different power-law mass
functions ($dN/dM \propto M^{-\alpha}$) with $\alpha$ = $-$1, 0, and +1 (solid black) having a minimum formation
mass of 1 $M_{Jup}$. Also shown for the $\alpha$ = 0 model
is the predicted number of brown dwarfs if a minimum formation mass of 5 $M_{Jup}$ (dashed black) or 10 $M_{Jup}$ (dotted black) is assumed. These
simulations are from \cite{burgasser2004,burgasser2007}.
Space densities using our full accounting of objects in the immediate Solar Neighborhood (Tables~\ref{TY_census_distances} 
and~\ref{space_density_numbers}) are shown by the heavy purple line.
\label{space_density1}}
\end{figure}

\begin{figure}
\epsscale{0.8}
\figurenum{15}
\plotone{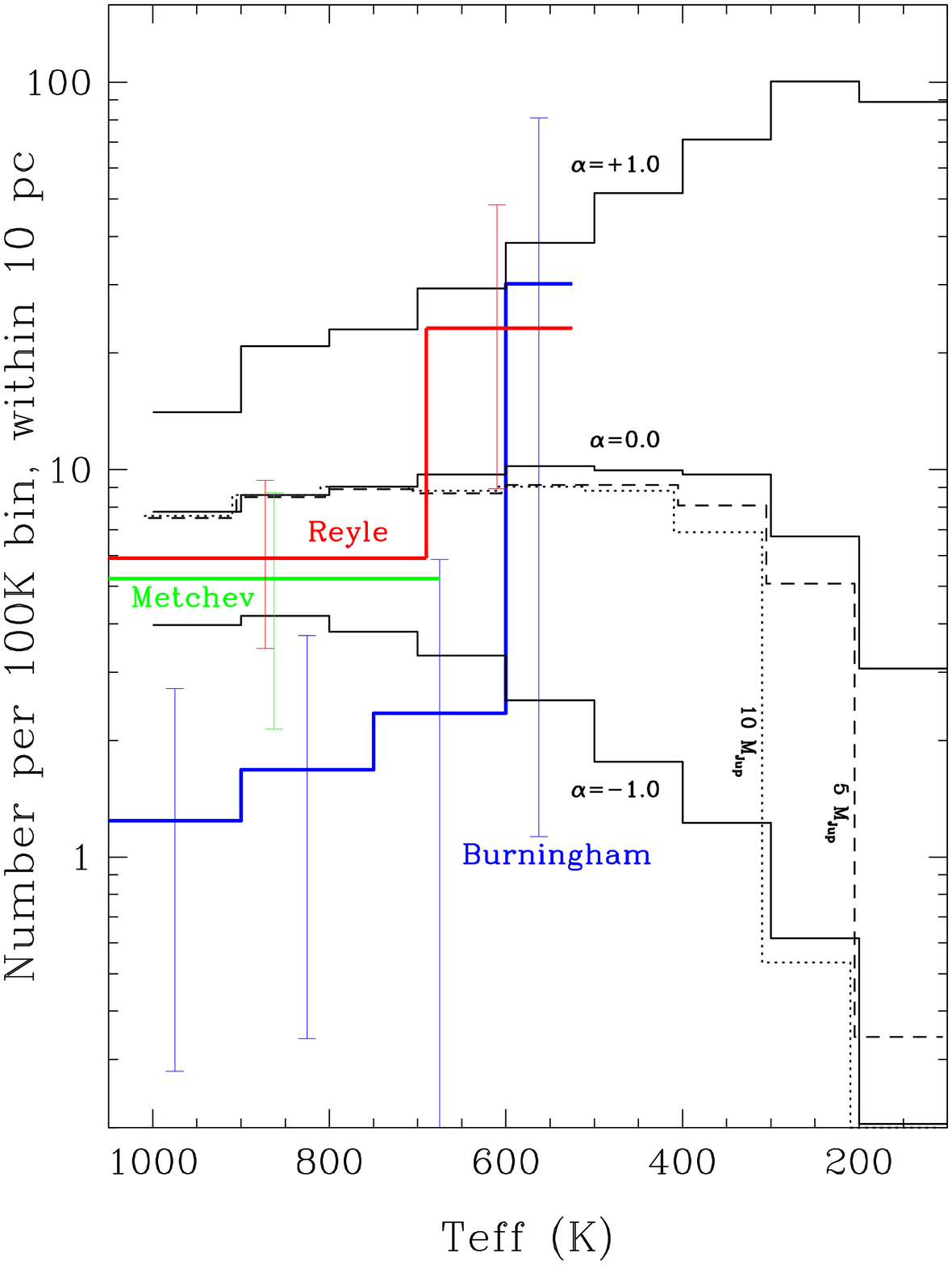}
\caption{Previous measurements of the space density of mid- to late-T dwarfs from \cite{metchev2008} (green),
\cite{reyle2010} (red), and \cite{burningham2010} (blue) overplotted on the same simulations from Figure~\ref{space_density1}.
\label{space_density2}}

\end{figure}

\end{document}